\documentclass[a4paper,12pt]{article}
\usepackage{textgreek}
\usepackage[utf8]{inputenc}
\usepackage{graphicx}
\usepackage{amssymb,amsmath,amsthm}
\usepackage{subfigure,subfloat}
\usepackage[margin=2.5cm]{geometry}
\usepackage{authblk}
\usepackage{url}

\graphicspath{{figs/}}

\title{Actin droplet machine}
\author[1]{Andrew Adamatzky}
\affil[1]{Unconventional Computing Laboratory, Department of Computer Science, University of the West of England, Bristol, UK}
\author[2]{Florian Huber}
\affil[2]{Netherlands eScience Center, Science Park 140, 1098 XG Amsterdam, The Netherlands}
\author[3]{J\"{o}rg Schnau{\ss}}
\affil[3]{Soft Matter Physics Division, Peter Debye Institute for Soft Matter Physics, Faculty of Physics and Earth Science, Leipzig University, Germany \& Fraunhofer Institute for Cell Therapy and Immunology (IZI), DNA Nanodevices Group, Leipzig, Germany}
\date{}

\date{}

\begin{document}

\maketitle

\begin{abstract}
\noindent
The actin droplet machine is a computer model of a three-dimensional network of actin bundles developed in a droplet of a physiological solution, which implements mappings of sets of binary strings. The actin bundle network is conductive to travelling excitations, i.e. impulses. The machine is interfaced with an arbitrary selected set of $k$ electrodes through which stimuli, binary strings of length $k$ represented by impulses generated on the electrodes, are applied and responses are recorded. The responses are recorded in a form of impulses and then converted to binary strings. The machine's state is a binary string of length $k$: if there is an impulse recorded on the $i$th electrode, there is a `1' in the $i$th position of the string, and `0' otherwise. We present a design of the machine and analyse its state transition graphs. We envisage that actin droplet machines could form an elementary processor of future massive parallel computers made from biopolymers.

\vspace{2mm}

\end{abstract}

\section{Introduction}

Actin is a  protein presented in forms of monomeric, globular actin (G-actin) and filamentous actin (F-actin)~\cite{straub1943actin, korn1982actin, szent2004early}. G-actin polymerises into filamentous actin forming a double helical structure~\cite{straub1950muscle,korn1987actin,huber2013advances}. The filaments can be further arranged into bundles by various different mechanisms such as crowding effects, cross-linking or counterion condensation~\cite{wang1984reorganization,verkhovsky1995myosin,bartles2000parallel,schnaussPRL2016,schnaussreview2016,Strehle2011,strehle2017,huber2012counterion,huber2015formation}. The bundles are conductive to travelling localisations --- defects, ionic waves, solitons~\cite{tuszynski1995ferroelectric,tuszynski2004results,tuszynski2004ionic, tuszynski2005molecular, priel2006ionic, tuszynski2005nonlinear,sataric2009nonlinear, sataric2010solitonic,sataric2011ionic, kavitha2017localized}. By interpreting presence or absence of a travelling localisation at a given site of the network at a given time step, we can implement a logical function. This approach was comprehensively developed and successfully tested on chemical systems in the framework of collision-based computing ~\cite{badamatzky,jakubowski2002computing,blair2002gated,toth2009simple,zhang,adamatzkyCBC}. 

Our approach --- computing with excitation waves propagating on overall `density' of the conductive material --- has previously been presented by us in~\cite{AdamatzkyHuberSchnauss}. As conductive material we looked at networks of actin bundles which were arranged by crowding effects without the need of additional accessory proteins~\cite{schnaussPRL2016,schnaussreview2016}. We demonstrated how to discover logical gates on a two-dimensional slice of the actin bundle network by representing Boolean inputs and outputs as spikes of the network activity. 

In the present paper we  develop a novel concept and computer modelling implementation of the actin network machine, which implements a 
mapping $F: \{0, 1\}^k \rightarrow \{0, 1\}^k$, where $k$ is a  number of electrodes, and `1' signifies a presence of an impulse on the electrode and `0' the absence. At a higher level, the machine acts as a finite state machine, at the lower level a structure of the mapping $F$ is determined by interactions of impulses propagating on the three-dimensional network of actin bundles.

We also offer an alternative to a numerical integration used in \cite{AdamatzkyHuberSchnauss}: an automaton model of a three-dimensional actin network. There is a substantial body of evidence confirming that automaton models are sufficient and appropriate discrete tools for modelling dynamics of spatially extended non-linear excitable media~\cite{markus1990isotropic,gerhardt1990cellular,weimar1992diffusion},  propagation~\cite{lechleiter1991spiral}, action potential~\cite{ye2005efficient,atienza2005probabilistic},
electrical pulses in the heart~\cite{saxberg1991cellular,dowle1997fast,siregar1998interactive}. A major advantage of automata is that they require less computational resources than typical numerical integration approaches. 

The paper is structured as follows. Our modelling approach is described in detail in Sect.~\ref{methods}. This includes a representation of a three-dimensional actin bundle network (Subsect.~\ref{actinnetwork}), a structure of an automaton model to simulate propagation of impulses on the actin bundle network (Subsect.~\ref{actinautomata}), and an interface with the actin network (Subsect.~\ref{interfacing}). In Sect.~\ref{frequencies} we analyse dependencies of a number of Boolean gates implemented in the network on an excitation threshold and refractory period. Thus, we justify the selection of these parameters for the construction of the actin machine. The actin droplet machine is designed and analysed in Sect.~\ref{machine}. Section~\ref{discussion}  discusses the results in a context of cytoskeleton computing and outlines directions for future research.

\section{Methods}
\label{methods}

  \begin{figure}[!tbp]
     \centering
     \includegraphics[width=0.8\textwidth]{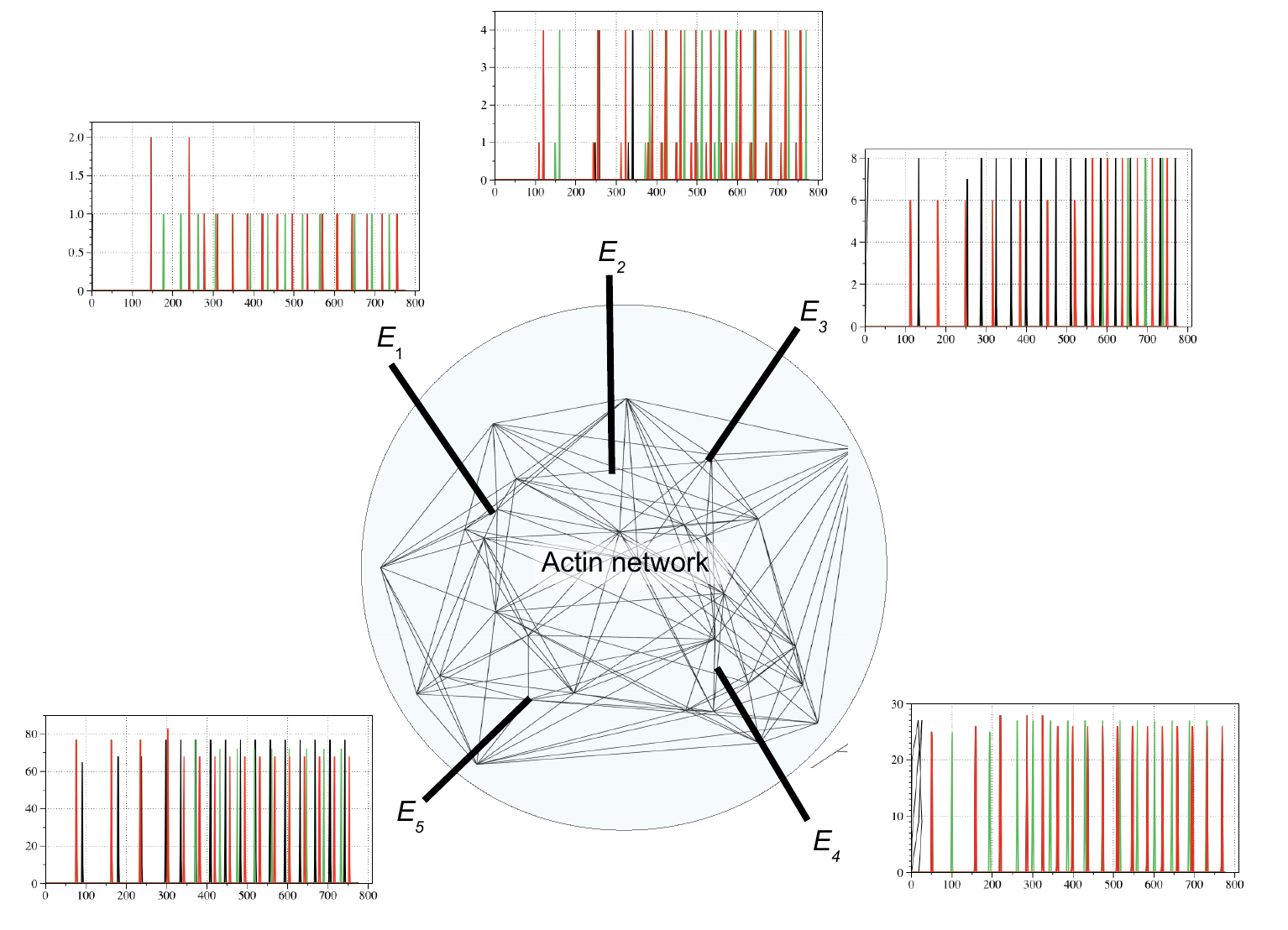}
     \caption{A scheme of a virtual experiment. The actin bundle network is shown as a three-dimensional Delaunay triangulation. Electrodes are shown by thick lines and labelled $E_1$ to $E_5$. Exemplary trains of spikes are shown near the electrodes.}
     \label{fig:overallidea}
 \end{figure}

The overall approach is the following: we simulate the actin bundle network using three-dimensional arrays of finite-state machines, cellular automata. We select several domains of the network and assign them as inputs and outputs.
We represent Boolean logic values with spikes of electrical activity, which are schematically represented as a virtual experiment in Fig.~\ref{fig:overallidea}. We stimulate the network with all possible configurations of input strings and record spikes on the outputs. Based on the mapping of configurations of input spikes to output spikes, we reconstruct logical functions implemented by the network.  In our design of the actin droplet machine we consider outputs recorded on all electrodes at a given time step as a binary string and then represent the actin droplet machine as a finite-state machine whose states are binary strings of a given length. 

\subsection{Three-dimensional actin network}
\label{actinnetwork}

 \begin{figure}[!tbp]
     \centering
\includegraphics[width=\textwidth]{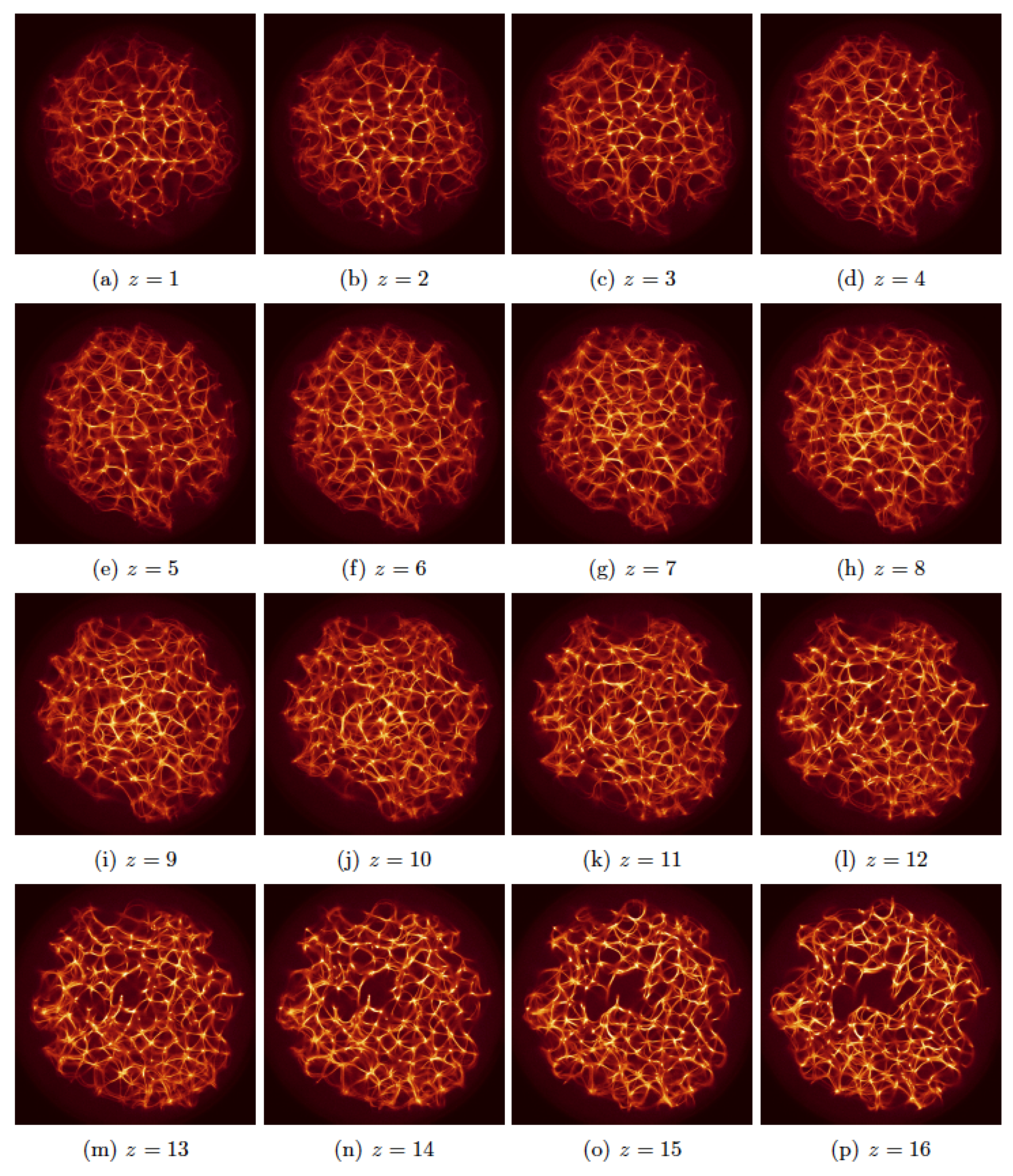}
     \caption{Exemplary $z$-slices of a three-dimensional actin bundle network reconstructed as described  in~\cite{huber2015formation}.}
     \label{fig:sampleslicesexperimental}
 \end{figure}

 \begin{figure}[!tbp]
     \centering
\includegraphics[width=\textwidth]{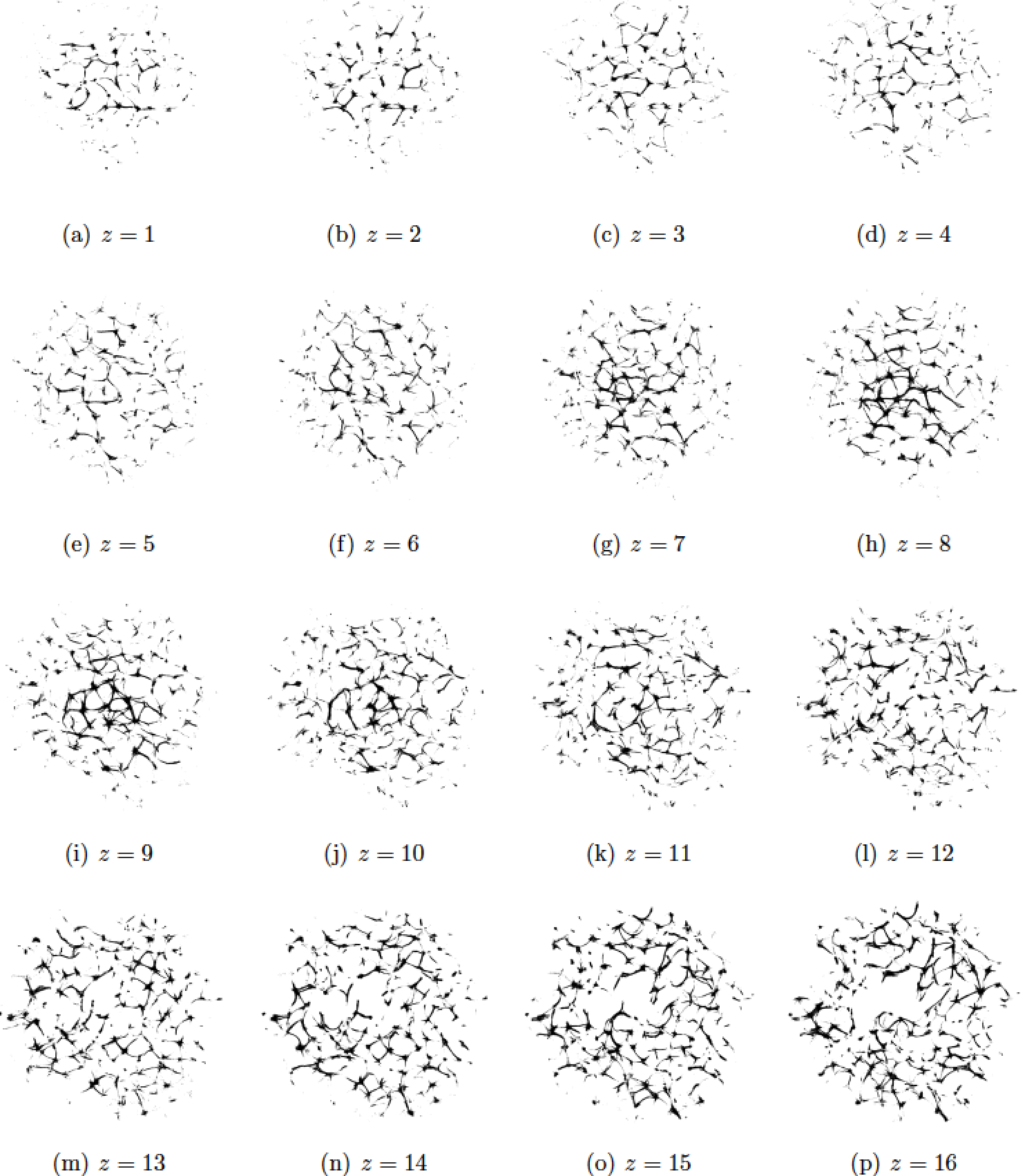}
     \caption{Exemplary $z$-slices of `conductive' geometries $C$ selected from the three-dimensional actin bundle network shown in  Fig.~\ref{fig:sampleslicesexperimental}, which were reconstructed as described  in~\cite{huber2015formation}.}
     \label{fig:sampleslices}
 \end{figure}
 
 As a template for our actin droplet machine we used an actual three-dimensional actin bundle network produced in laboratory experiments with purified proteins (Fig.~\ref{fig:sampleslicesexperimental}). The underlying experimental method was shown to reliably produce regularly spaced bundle networks from homogeneous filament solutions inside small isolated droplets in the absence of molecular motor-driven processes or other accessory proteins~\cite{huber2015formation}. These structures effectively form very stable and long-living three-dimensional networks, which can be readily imaged with confocal microscopy resulting in stacks of optical two-dimensional slices (Fig.~\ref{fig:sampleslicesexperimental}).  Dimensions of the network are the following: size along $x$ coordinate is 225~$\mu$m (width), along $y$ coordinate is 222~$\mu$m (height), along $x$ coordinate is 112~$\mu$m (depth), voxel width is 0.22~$\mu$m, height 0.22~$\mu$m and depth 4~$\mu$m.
 
 Original image:  
 $A_z=(a_{ijz})_{1 \leq i,j \leq n, 1 \leq z \leq m}$, 
 $a_{ijz} \in \{ r_{ijz}, g_{ijz}, b_{ijz} \}$, 
 where $n=1024$, $m=30$, $r_{ijz}, g_{ijz}, b_{ijz}$ are RGB values of the element at $ijz$, $1 \leq r_{ijz}, g_{ijz}, b_{ijz} \leq 255$ was converted to a conductive matrix $\mathbf{C}=(c_{ijz})_{1 \leq i,j \leq n, 1 \leq z \leq m}$ as follows: $c_{ijz}=1$  if $r_{ijz}>40$, $g_{ijz}>19$ and $b_{ijz}>19$. The conductive matrices are shown in Fig.~\ref{fig:sampleslices}. The 3D conductive matrix is compressed along $z$-axis to reduce consumption of computational resources, scenario of the non-compressed matrix will be considered in future papers.

 \subsection{Automaton model}
 \label{actinautomata}
 
To model activity of an actin bundle network we represent it as an automaton 
${\mathcal A}=\langle \mathbf{C}, \mathbf{Q}, r, h, \theta, \delta   \rangle$.
$\mathbf{C} \subset \mathbf{Z}^3$ is a set of voxels, or a conductive matrix
$\mathbf{C}$ defined in Sect.~\ref{actinnetwork}. 
Each voxel $p \in \mathbf{C}$ takes states from the set $\mathbf{Q}=\{ \star, \bullet, \circ \}$, excited ($\star$), refractory ($\bullet$), resting ($\circ$) and is complemented by a counter $h_p$ to handle the temporal decay of the refractory state. Following discrete time steps, each voxel $p$ updates its state depending on its current state and the states of its neighbourhood 
$u(p)=\{ q \in \mathbf{C}: d(p,q) \leq r  \}$, where $d(p,q)$ is an Euclidean distance between voxels $p$ and $q$; $r \in {\bf N}$ is a neighbourhood radius. $\theta \in {\bf N}$ is an excitation threshold and $\delta \in {\bf N}$ is refractory delay. All voxels update their states in parallel and by the same rule:
$$
p^{t+1}
\begin{cases}
\star, \text{ if } (p^t=\circ) \text{ and } (\sigma(p)^t > \theta) \\
\bullet, \text{ if } (p^t=\circ) \text{ or } ((p^t=\bullet) \text{ and } (h_p^t>0))\\
\circ, \text{ otherwise}
\end{cases}
$$
$$
h_p^{t+1}=
\begin{cases}
\delta, \text{ if } (p^{t+1}=\bullet) \text{ and } (p^t=\star) \\
h_p^{t}-1, \text{ if } (p^{t+1}=\bullet) \text{ and } (h_p^t>0)\\
0, \text{ otherwise}.
\end{cases}
$$
Every resting ($\circ$) voxel of $\mathbf{C}$ excites ($\star$) at the moment $t+1$ if a number of its excited neighbours at the moment $t$,  $\sigma(p)^t=|\{q \in u(p): q^t=\star\}|$, exceeds a threshold $\theta$. An excited voxel $p^t=\star$ takes the refractory state $\bullet$ at the next time step $t+1$ and at the same moment a counter of refractory state $h_p$ is set to the refractory delay $\delta$. The counter is decremented, $h_p^{t+1}=h_p^t-1$ at each iteration until it becomes 0. When the counter $h_p$ becomes zero the voxel $p$ returns to the resting state $\circ$. For all results shown in this manuscript, the neighbourhood radius was set to $r=3$. Choices of $\theta$ and $\delta$ are considered in Sect.~\ref{frequencies}.

\subsection{Interfacing with the network}
\label{interfacing}

 To stimulate the network and to record activity of the network we assigned several domains of $\mathbf{C}$ as electrodes. 
 We calculated a potential $p^t_x$ at an electrode location $c \in \mathbf{C}$ as $p_c = |{z: d(c,z)<r_e \text{ and } z^t=+}|$, where $d(c,z)$ is an Euclidean distance between sites $x$ and $z$ in 3D space. We have chosen an electrode radius of $r_e=4$ voxels and conducted two families of experiments with two configurations of electrodes. 

\begin{table}
\parbox{.45\linewidth}{
\centering
  \begin{tabular}{c|ccc}
$e$ &      $i$  &    $j$   &   $z$ \\ \hline
1	&	369	&	567	&	6	\\
2	&	509	&	580	&	10	\\
3	&	631	&	590	&	10	\\
4	&	382	&	322	&	12	\\
5	&	533	&	331	&	23	\\
6	&	626	&	463	&	7	\\
7	&	358	&	676	&	22	\\
8	&	369	&	424	&	7	\\
9	&	572	&	691	&	17	\\
10	&	705	&	394	&	17	\\
    \end{tabular}
    \label{tab:e1}
\caption{Coordinates of electrodes in experiments family $\mathcal{E}_1$.}
}
\hfill
\parbox{.45\linewidth}{
\centering
\begin{tabular}{c|ccc}
$e$ &      $i$  &    $j$   &   $z$ \\ \hline
1	&	369	&	567	&	6	\\
2	&	509	&	580	&	10	\\
3	&	631	&	590	&	10	\\
4	&	382	&	322	&	12	\\
5	&	533	&	331	&	23	\\
6	&	369	&	424	&	7	\\
7	&	572	&	691	&	17	\\
8	&	705	&	394	&	17	\\
    \end{tabular}
    \label{tab:e2}
\caption{Coordinates of electrodes in experiments family $\mathcal{E}_2$.}
}
\end{table}

 \begin{figure}
    \centering
    \subfigure[]{\includegraphics[width=0.49\textwidth]{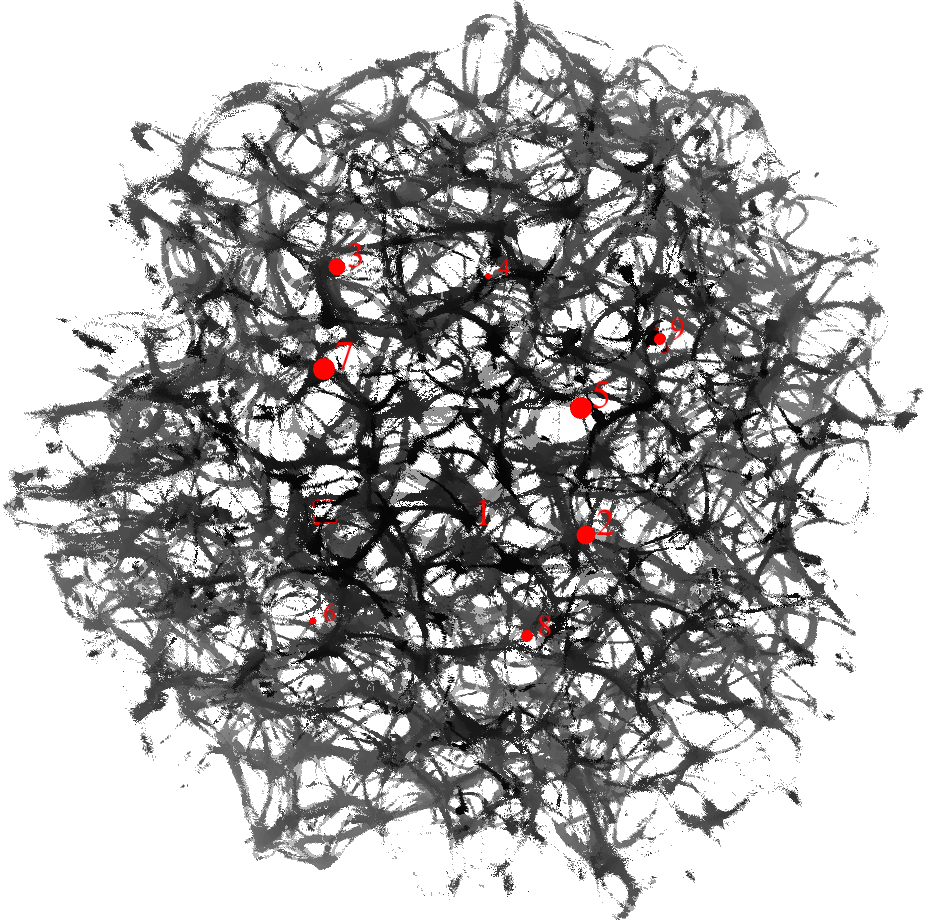}\label{electrodesE1}}
    \subfigure[]{\includegraphics[width=0.49\textwidth]{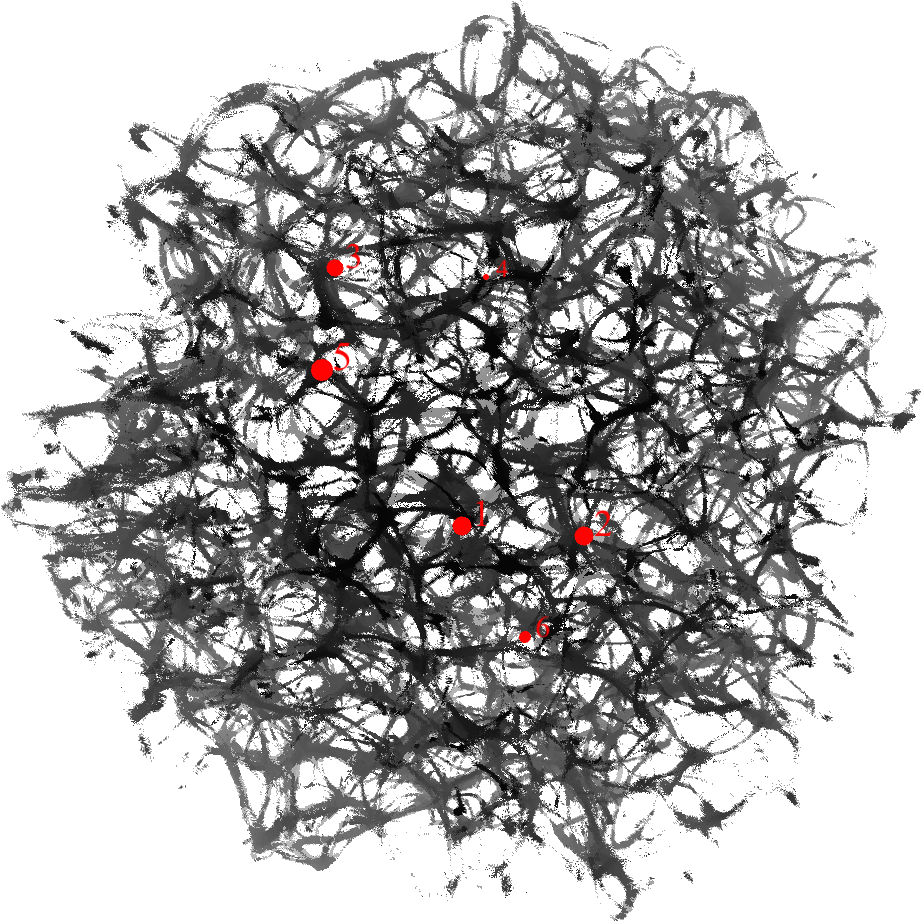}\label{electrodesE2}}
    \caption{Configurations of electrodes in the three-dimensional network of actin bundles used in (a)~$\mathcal{E}_1$ and (b)~$\mathcal{E}_2$. Depth of the network is shown by level of grey. Sizes of the electrodes are shown in perspective. }
    \label{fig:locationelectrodes}
\end{figure}

In the first family of experiments $\mathcal{E}_1$
we studied frequencies of two-input-one-output Boolean functions implementable in the network. We used ten electrodes, their coordinates are listed in Tab.~\ref{tab:e1} and a configuration is shown in Fig.~\ref{electrodesE1}.  Electrodes $E_0$ representing input $x$ and $E_9$ representing input $y$ are the input electrodes, all others are output electrodes representing outputs $z_1, \ldots, z_8$.  Results are presented in Sect.~\ref{frequencies}. In the second family of experiments $\mathcal{E}_2$ we used six electrodes (Tab.~\ref{tab:e2} and Fig.~\ref{electrodesE2}). All electrodes were considered as inputs during stimulation and outputs during recording of the network activity.

\begin{figure}[!tbp]
    \centering
\subfigure[$t=13$]{\includegraphics[width=0.49\textwidth]{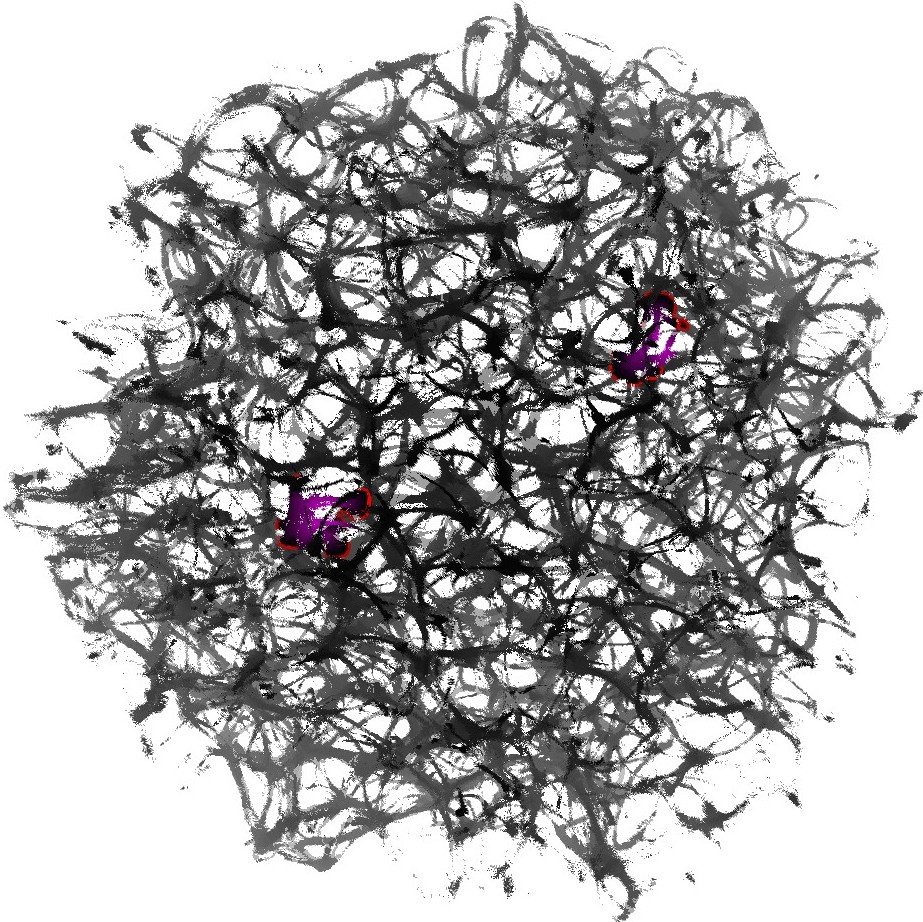}\label{snapshots:t13}}
\subfigure[$t=50$]{\includegraphics[width=0.49\textwidth]{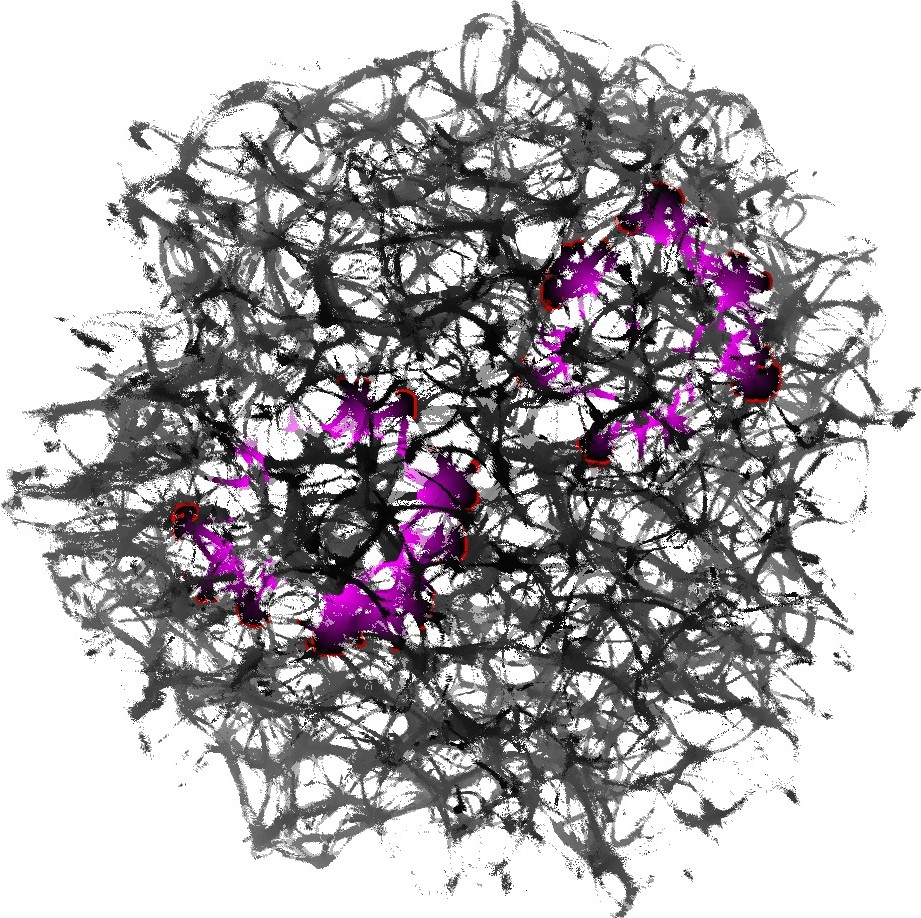}\label{snapshots:t50}}
\subfigure[$t=200$]{\includegraphics[width=0.49\textwidth]{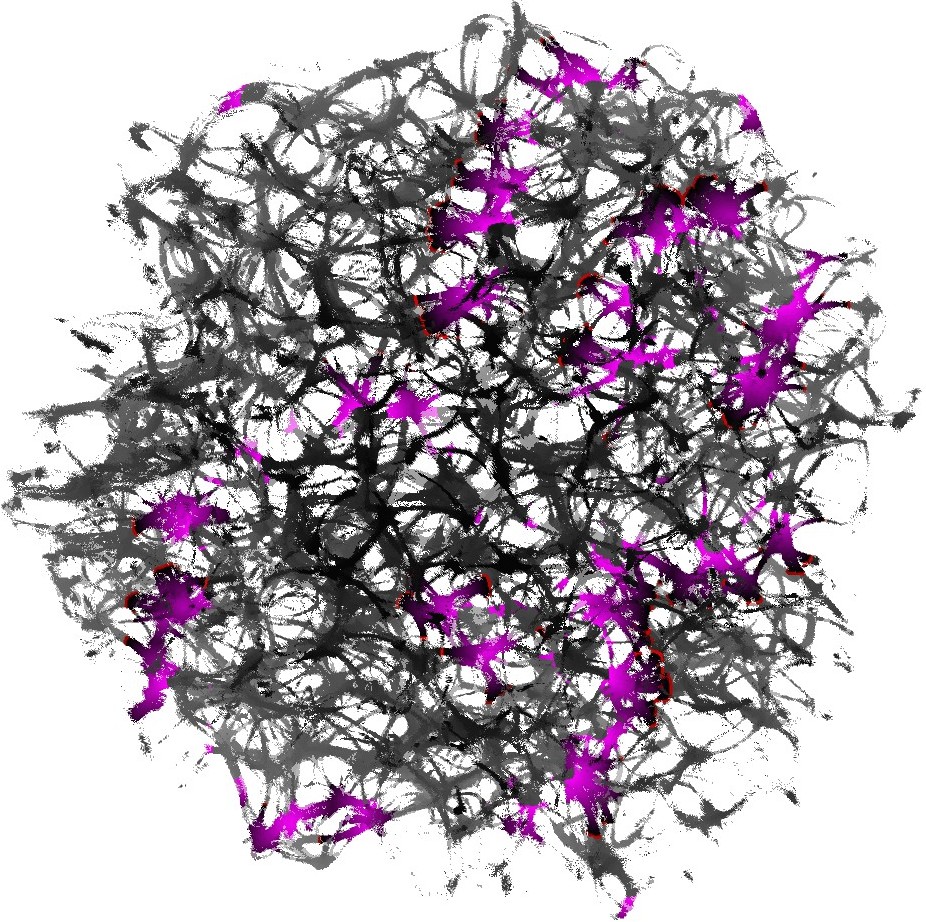}\label{snapshots:t200}}
\subfigure[$t=500$]{\includegraphics[width=0.49\textwidth]{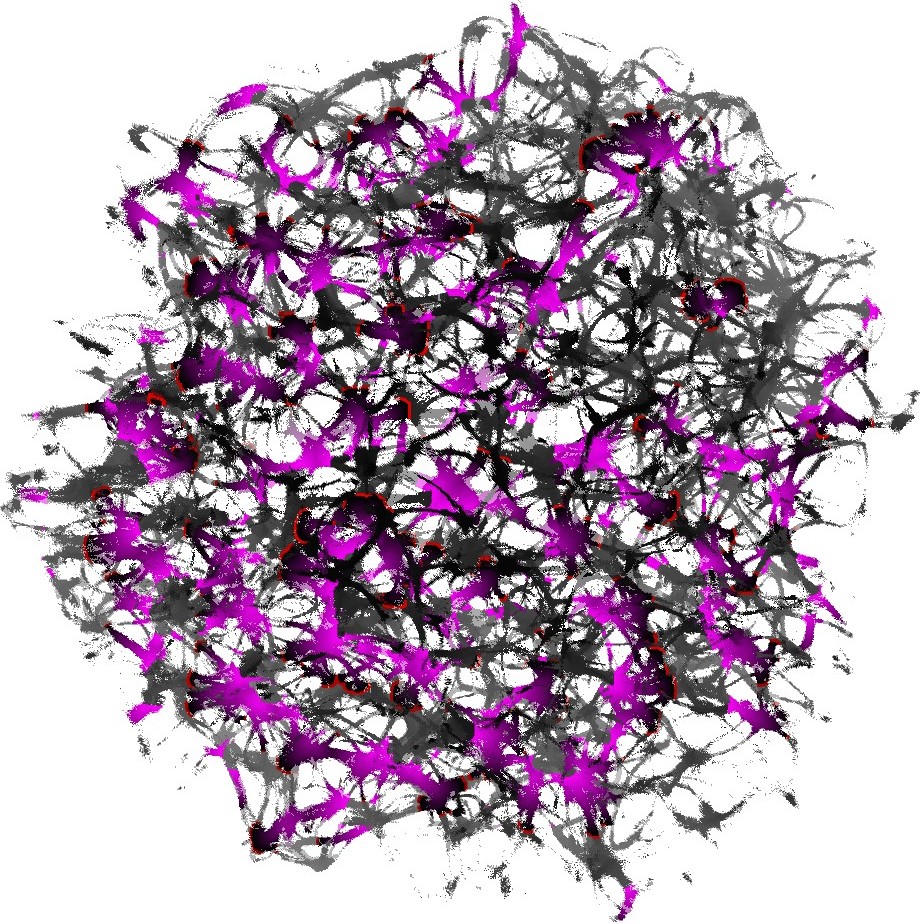}\label{snapshots:t500}}
    \caption{Snapshots of excitation dynamics on the network. The excitation wave front is red and the refractory tail is magenta. The excitation threshold is $\theta=7$ and the refractory delay  is $\delta=20$.}
    \label{fig:snapshots}
\end{figure}

Exemplary snapshots of excitation dynamics on the network are shown in Fig.~\ref{fig:snapshots}. Domains corresponding to the two electrodes $e_0$ and $e_9$ (Tab.~\ref{tab:e1} and Fig.~\ref{electrodesE1}) have been excited (Fig.~\ref{snapshots:t13}). The excitation wave fronts propagates away from $e_0$ and $e_9$ (Fig.~\ref{snapshots:t50}). The fronts traverse the whole breadth of the network  (Fig.~\ref{snapshots:t200}). Due to the presence of circular conductive paths in the network, the repetitive patterns of activity emerge (Fig.~\ref{snapshots:t500}). Videos of the experiments can be found in \url{http://doi.org/10.5281/zenodo.2649293}.

\section{Frequencies of gates}
\label{frequencies}

\begin{figure}[!tbp]
    \centering
\subfigure[]{\includegraphics[width=0.49\textwidth]{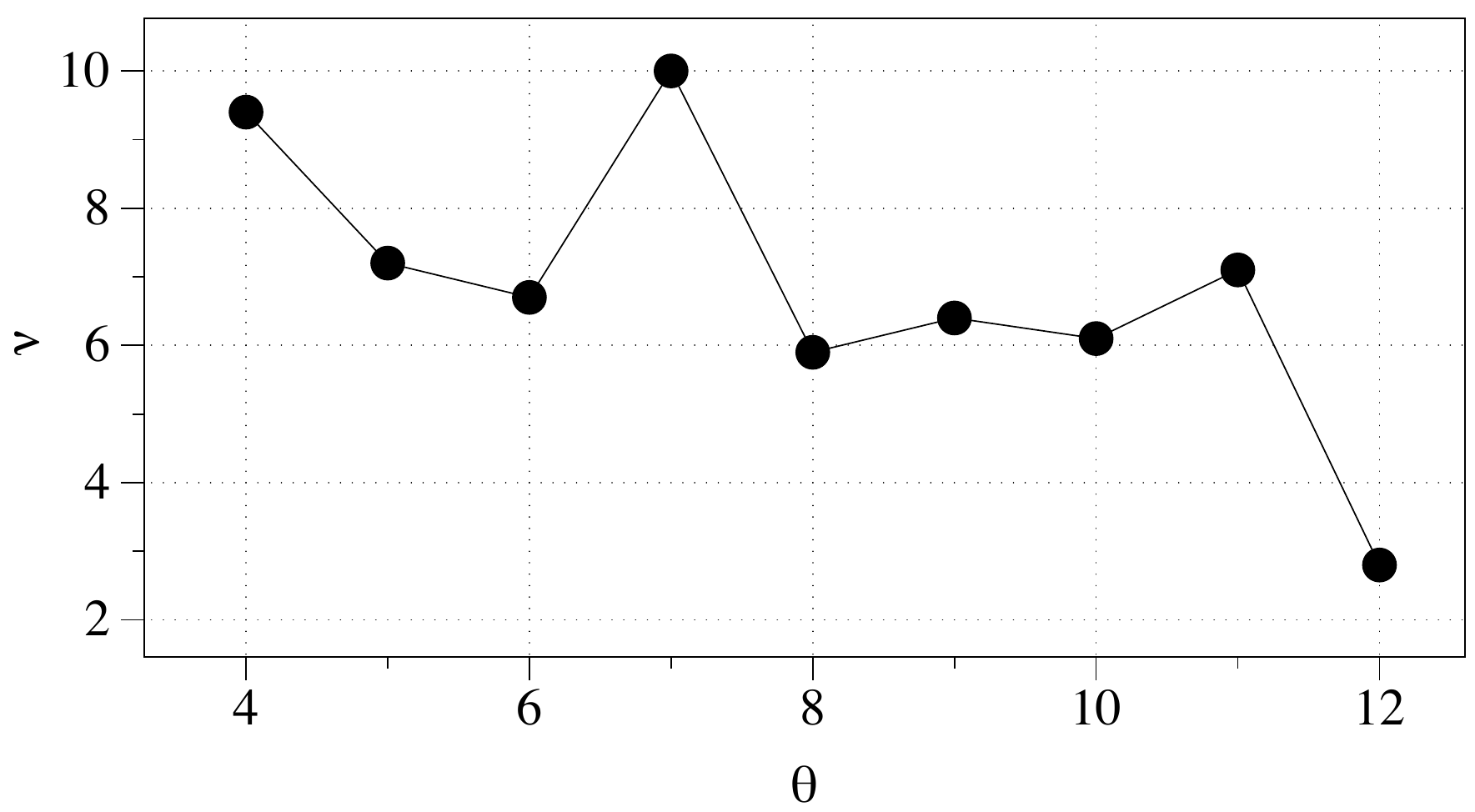}\label{fig:allgatestheta}}
\subfigure[]{\includegraphics[width=0.49\textwidth]{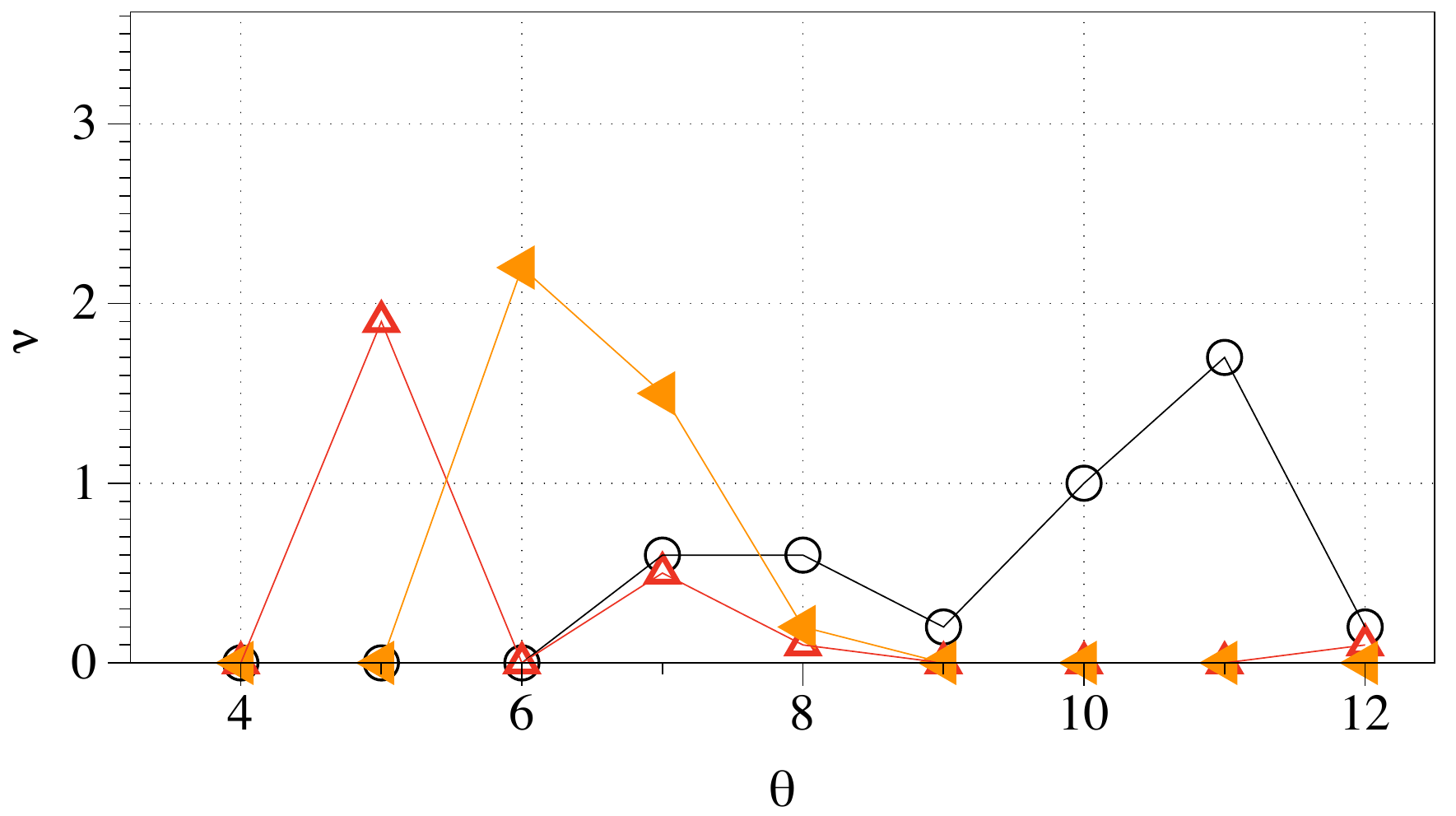}\label{fig:or_xor_and_theta}}
\subfigure[]{\includegraphics[width=0.49\textwidth]{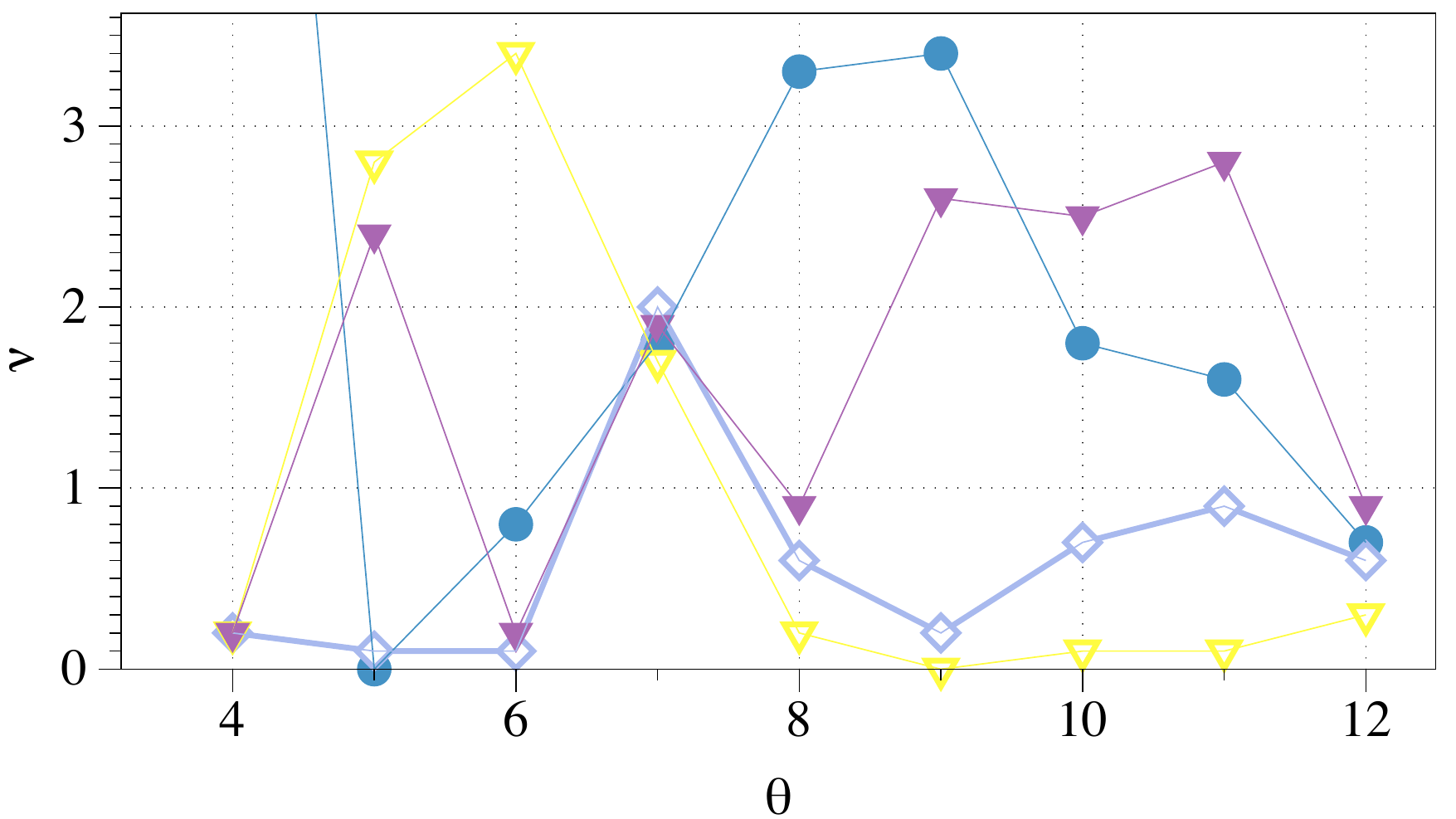}\label{fig:AN_NA_Sx_Sy}}
\subfigure[]{\includegraphics[width=0.49\textwidth]{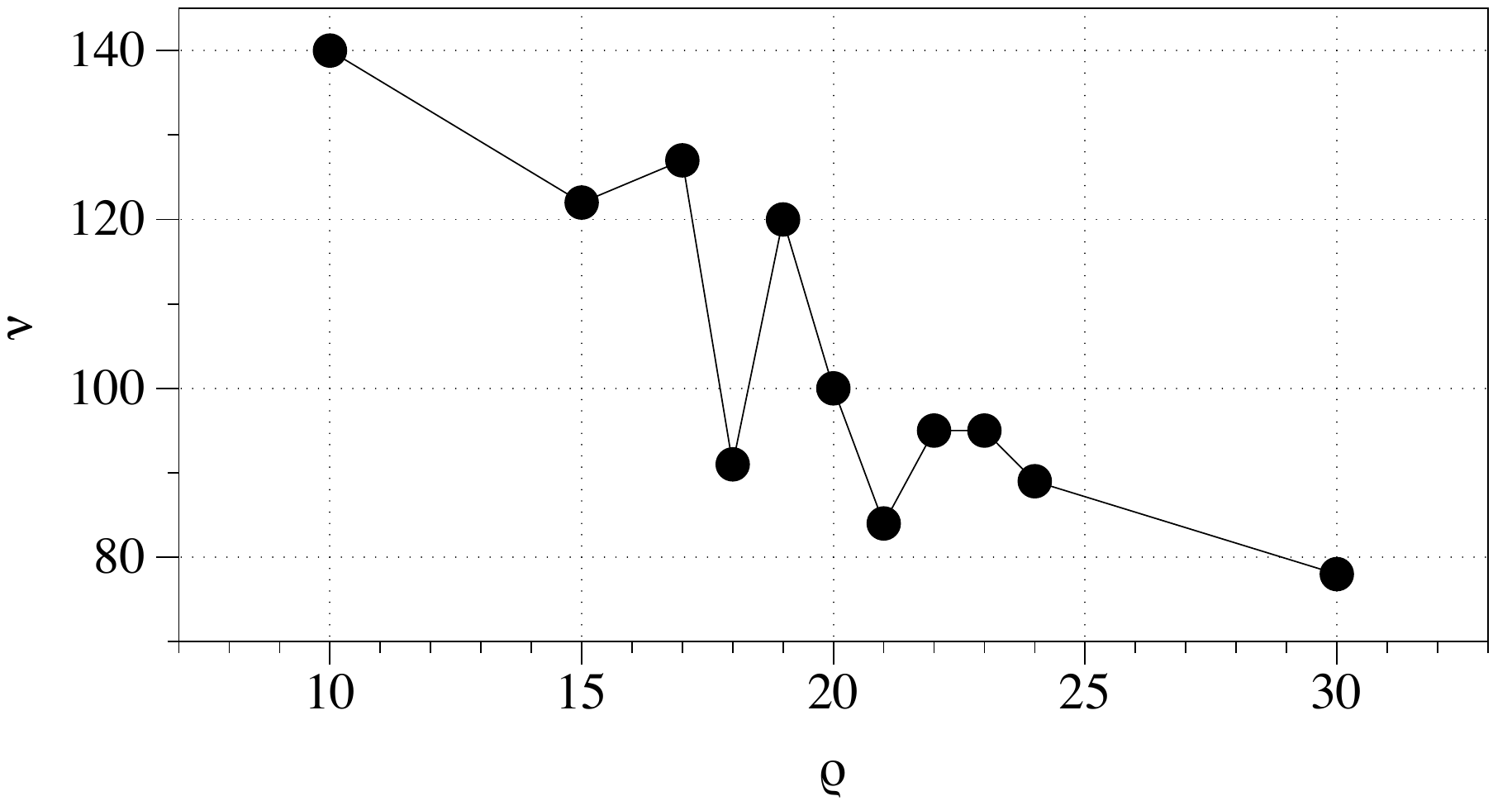}\label{fig:NumGates_vs_RefD}}
\subfigure[]{\includegraphics[width=0.49\textwidth]{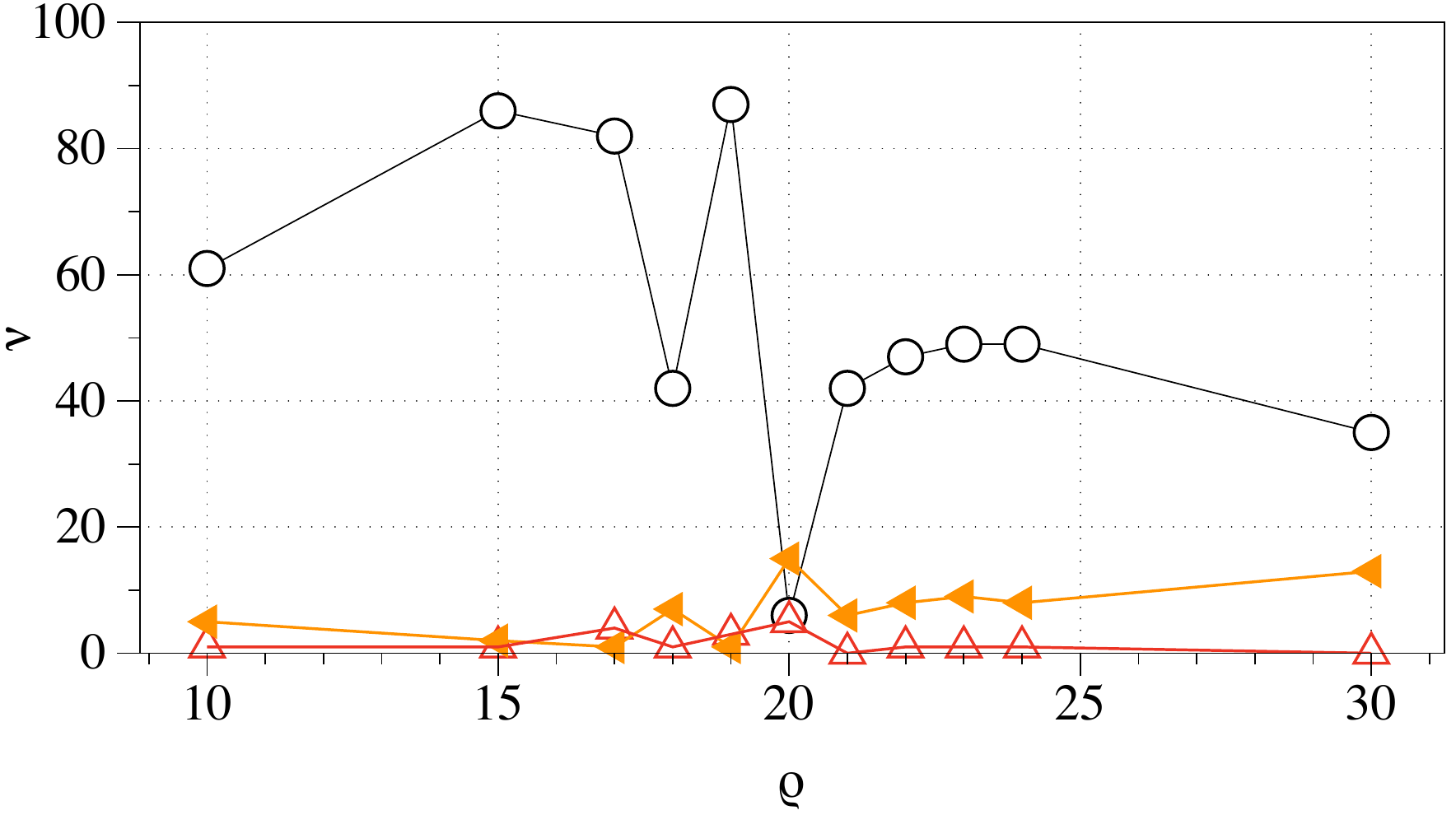}\label{fig:RefD_vs_OR_AND_XOR}}
    \caption{An average number $\nu$ of gates realisable on each of the electrodes $e_1, \ldots, e_8$ depends on threshold $\theta$ of excitation when the refractory delay $\delta$ is fixed to 20 (abc) and on refractory delays $\delta$ when the threshold $\theta$ is fixed to 7 (def).   
    (a)~Number of gates $\nu$ versus threshold $\theta$, $\delta=20$.
    (b)~Number of {\sc or} (black circle), {\sc and} (orange solid triangle) and {\sc xor} (red blank triangle) gates, $\delta=20$.
    (c)~Number of {\sc not-and} (yellow blank triangle), {\sc and-not} (magenta solid triangle), {\sc select}($x$) (cyan blank rhombus), {\sc select}($y$) (light blue disc), $\delta=20$.
    (d)~Number of gates $\nu$ versus delay $\delta$, $\theta=7$.
    (e)~Number of {\sc or} (black circle), {\sc 
    and} (orange solid triangle) and {\sc xor} 
    (red blank triangle) gates, $\theta=7$.}
    \label{fig:my_label}
\end{figure}

To map dynamics of the network onto sets of gates, we undertook the following trials of stimulation  
\begin{enumerate}
    \item fixed refractory delay $\delta=20$ and excitation threshold $\theta=4,5,\ldots,12$, 
    \item fixed excitation threshold $\theta=7$, and refractory delay $\delta=10, 15, 17, \ldots, 24, 30$. 
\end{enumerate}

For each combination $(\rho, \theta)$ we counted numbers of gates {\sc or}, {\sc and}, {\sc xor}, {\sc not-and}, {\sc and-not} and {\sc select}. We found that in overall a total number of gates $\nu(\theta)$ realised by the network decreases with increase of $\theta$ (Fig.~\ref{fig:allgatestheta}). The function $\nu(\theta)$ is non-linear and  could be adequately described by a five degree polynomial. The function reaches its maximal value at $\theta=7$ (Fig.~\ref{fig:allgatestheta}).  {\sc or} gates are most commonly realised at $\theta=11$, {\sc and} gates at $\theta=6$ and $\sc xor$ gates at $\theta=5$ as well as $\theta=7$ (Fig.~\ref{fig:or_xor_and_theta}. A number of {\sc and-not} gates implemented by the network reaches its highest value at $\theta=6$ then drops sharply after $\theta_8$ (Fig.~\ref{fig:AN_NA_Sx_Sy}). {\sc not-and} gates are more common at $\theta=5, 7, 9, 11$, while {\sc select}$(x)$ has its peak at $\theta=7$ and {\sc select}$(y)$ at $\theta=8, 9$ (Fig.~\ref{fig:AN_NA_Sx_Sy}). A total number of gates realised in the network with the excitability threshold fixed to $\theta=7$ decreases with the increase of $\delta$. Oscillations of $\nu(\delta)$ are visible at $15 \leq \delta \leq 25$ (Fig.~\ref{fig:NumGates_vs_RefD}). The three highest values of $\nu(\delta)$ are achieved at $\delta=10, 17$ and $20$. Let us look now at the dependence of the numbers of {\sc or}, {\sc and} and {\sc xor} gates of the refractory delay $\delta$ in Fig.~\ref{fig:RefD_vs_OR_AND_XOR}. The number of {\sc or} gates increases with $\delta$ increasing from 10 to 15, but then drops substantially at $\delta=18$ to reach its maximum at $\delta=19$. Numbers of gates {\sc and} and {\sc xor} behave similarly to each other. They both have a pronounced peak at $\delta=20$ (Fig.~\ref{fig:RefD_vs_OR_AND_XOR}). Thus, to maximise a number of logical gates produced and their diversity we selected $\theta=7$ and $\delta=20$ for our construction of the actin droplet machine.

\section{Actin droplet machine}
\label{machine}

An actin droplet machine is defined as a tuple 
${\mathcal M}=\langle  \mathcal{A}, k, \mathbf{E}, \mathbf{S}, F    \rangle$, where $\mathcal{A}$ is an actin network automaton, defined in Sect.~\ref{actinautomata}, 
$k$ is a number of electrodes, 
$\mathbf{E}$ is a configuration of electrodes, 
$\mathbf{S}=\{0, 1\}^k$, 
$F$ is a state-transition function $F: \mathbf{S} \rightarrow \mathbf{S}$ that implements a mapping between sets of all possible configurations of binary strings of length $k$. In the experiments reported here $k=6$.

\begin{figure}[!tbp]
    \centering
    \includegraphics[width=0.9\textwidth]{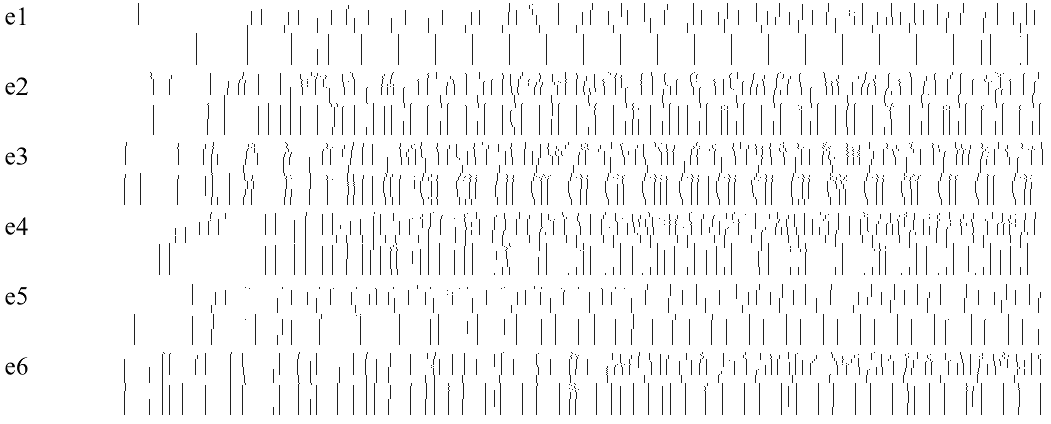}
    \caption{All spikes recorded at each electrode for input binary strings from 1 to 63. The representation is implemented as follows. We stimulate the $\mathcal{M}$ with strings from $\{ 0, 1 \}^6$ and represent a spike detected at time $t$ by  a black pixel at position $t$ along horizontal axis. A plot of each electrode $e_i$ represents a binary matrix 
    $\mathbf{S}=(s_{zt})$, where $1 \leq z \leq 63$ and $1 \leq t \leq 1000$: $s_{zt}=1$ if the input configuration was $z$  and a spike was detected at moment $t$, and $s_{zt}=1$ otherwise. }
    \label{fig:allspikes} 
\end{figure}

\begin{figure}[!tbp]
    \centering
    \subfigure[${\bf I}=5$]{
        \includegraphics[width=0.48\textwidth,trim={2.4cm 2.4cm 2.4cm 2.4cm},clip]{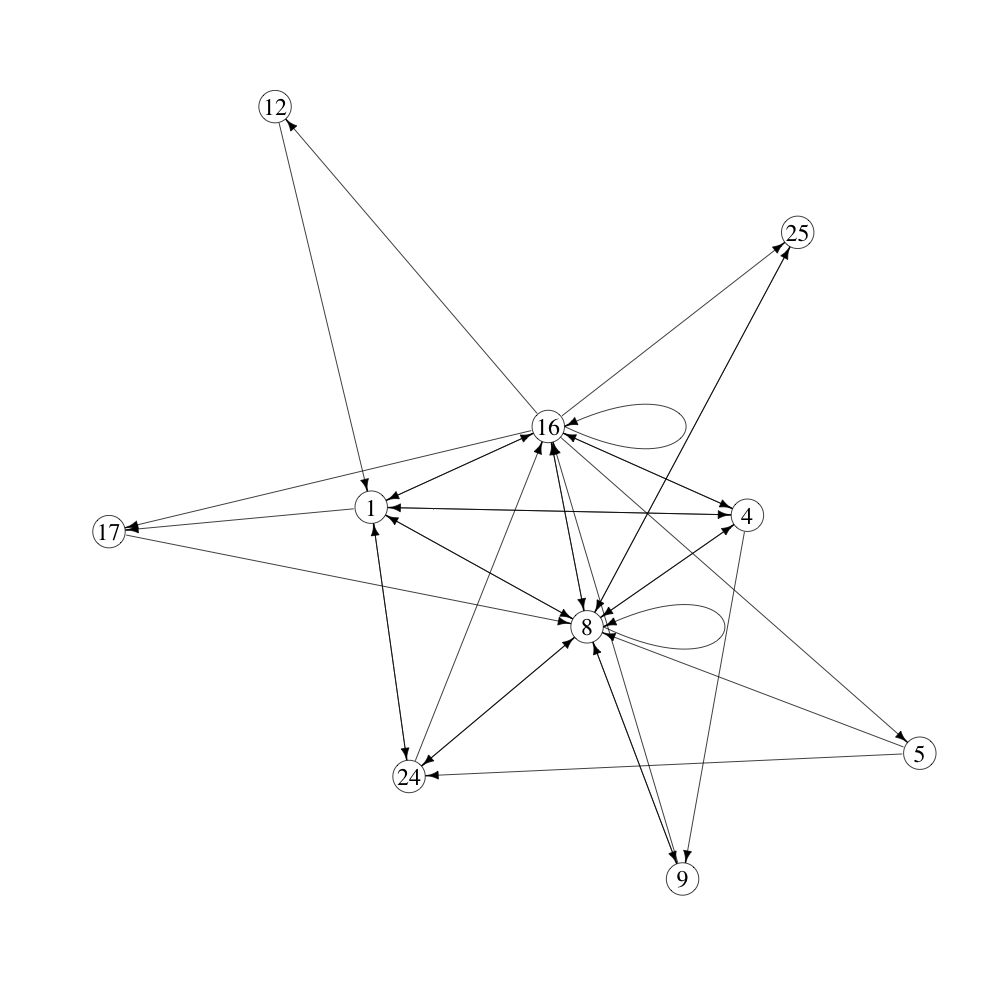}}
\subfigure[${\bf I}=15$]{
    \includegraphics[width=0.48\textwidth,trim={2.4cm 2.4cm 2.4cm 2.4cm},clip]{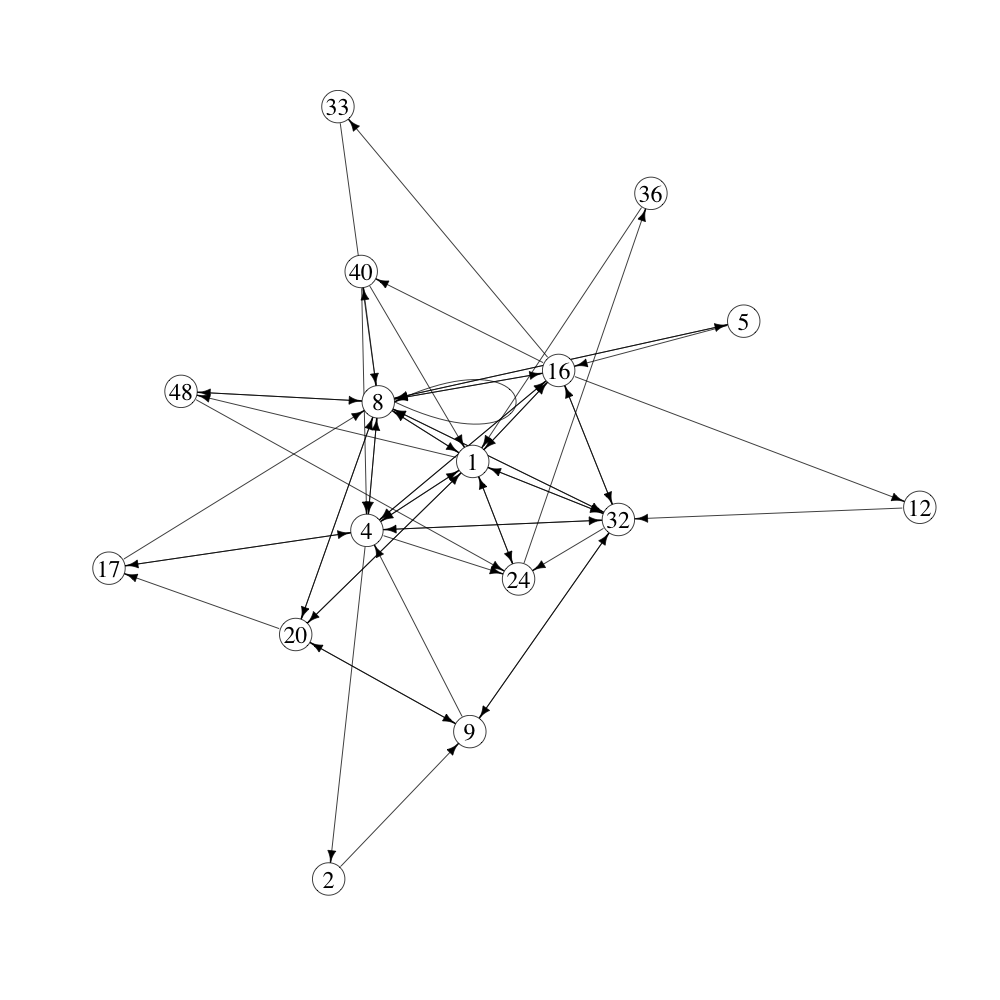}}
\subfigure[${\bf I}=31$]{
\includegraphics[width=0.48\textwidth,trim={2.4cm 2.4cm 2.4cm 2.4cm},clip]{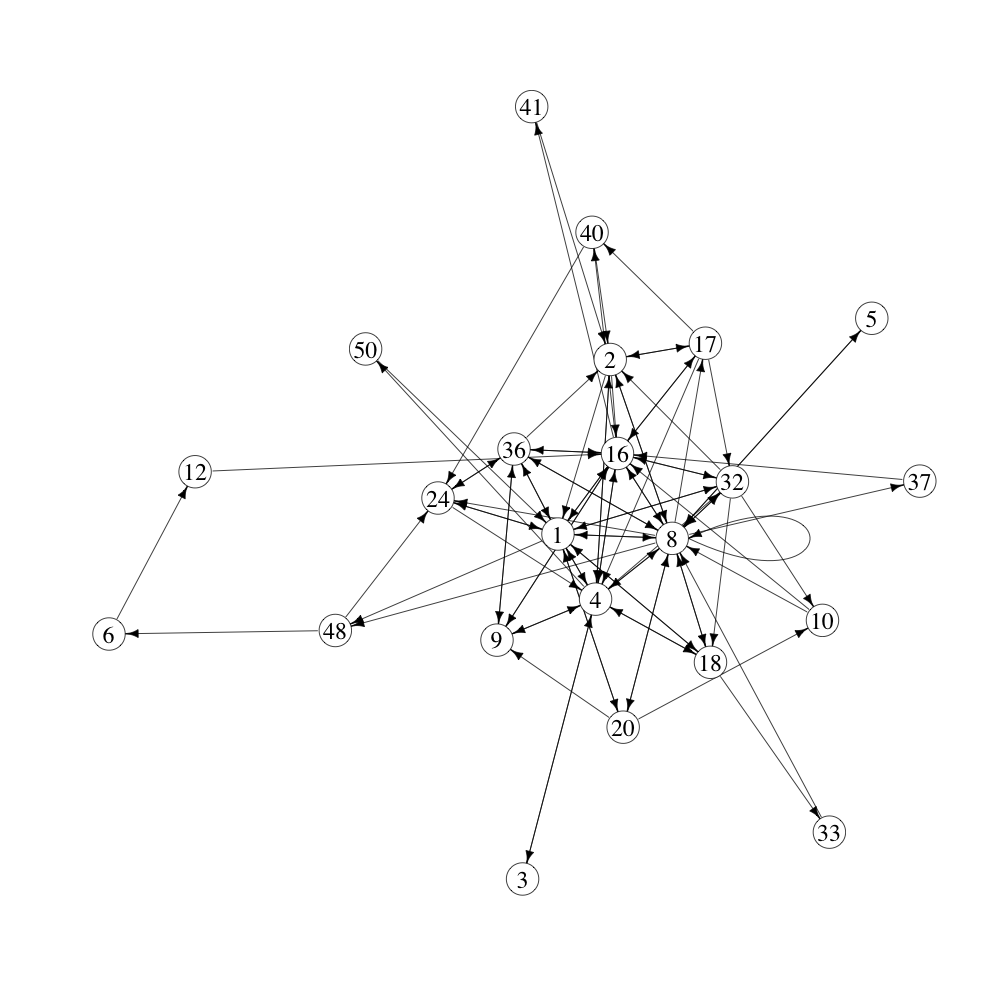} }  
\subfigure[${\bf I}=63$]{
\includegraphics[width=0.48\textwidth,trim={2.4cm 2.4cm 2.4cm 2.4cm},clip]{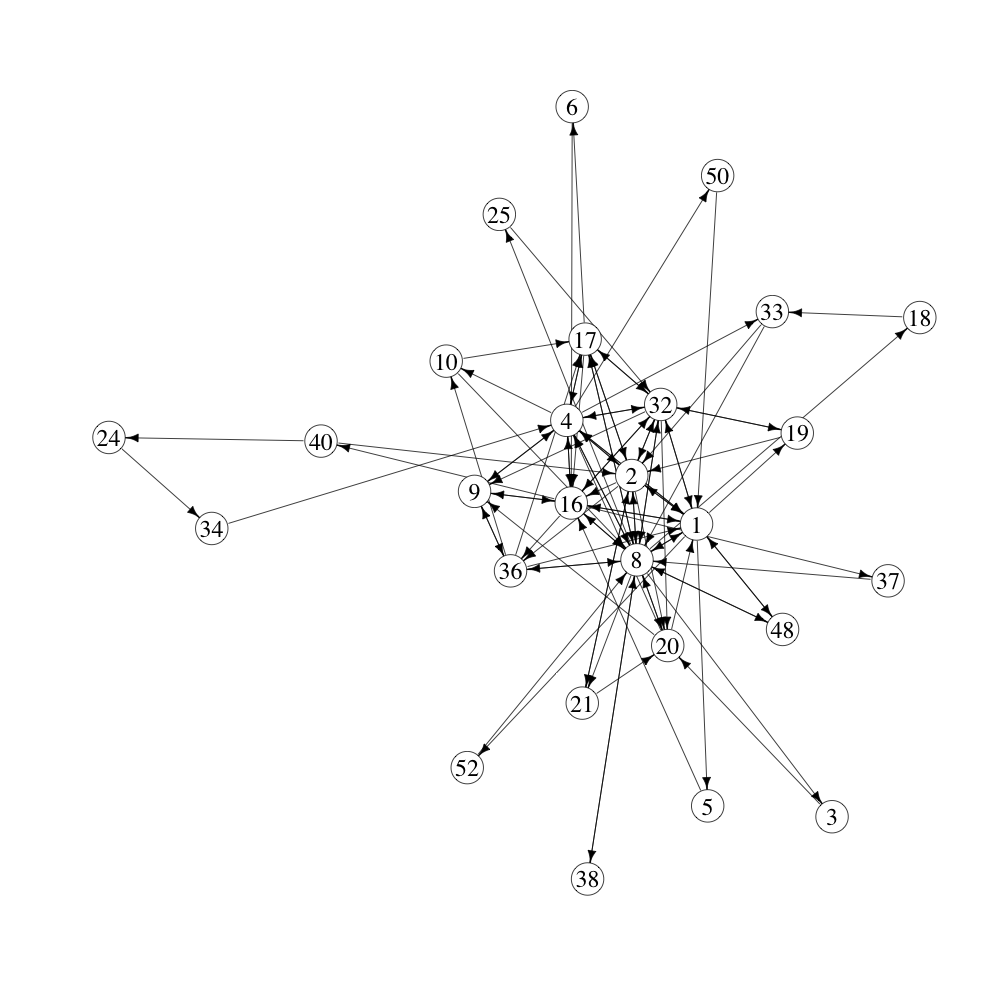}}
    \caption{State transitions of machine $\mathcal{M}$ for selected inputs $I$. A node is a decimal encoding of the $\mathcal M$ state $(e^t_0 \ldots e^t_5)$.}
    \label{fig:stateTransition}
\end{figure}

In our experiments we have chosen six electrodes, their locations are shown in Fig.~\ref{electrodesE2} and exact coordinates in Tab,~\ref{tab:e2}. Thus, $F: \{0,1\}^6 \rightarrow \{0,1\}^6$ and the machine $\mathcal M$ has 64 states. We represent the inputs and the machine states  in decimal encoding. Spikes detected in  response to every input from $\{0,1\}^6$ are shown in Fig.~\ref{fig:allspikes}.

\begin{figure}[!tbp]
    \centering
    \subfigure[]{
        \includegraphics[width=0.99\textwidth]{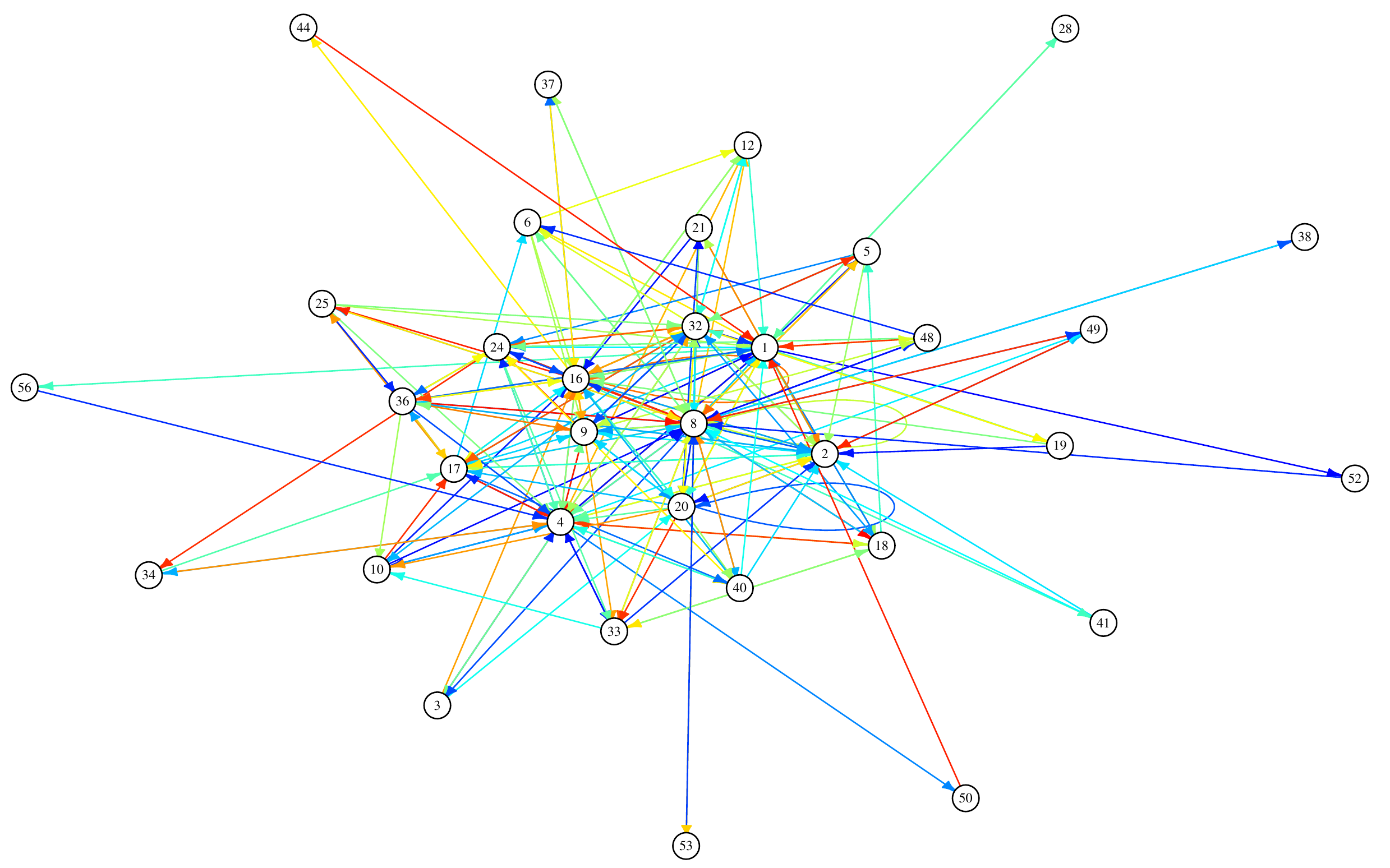}\label{fig:globalGraph}}
    \subfigure[]{
        \includegraphics[width=0.7\textwidth]{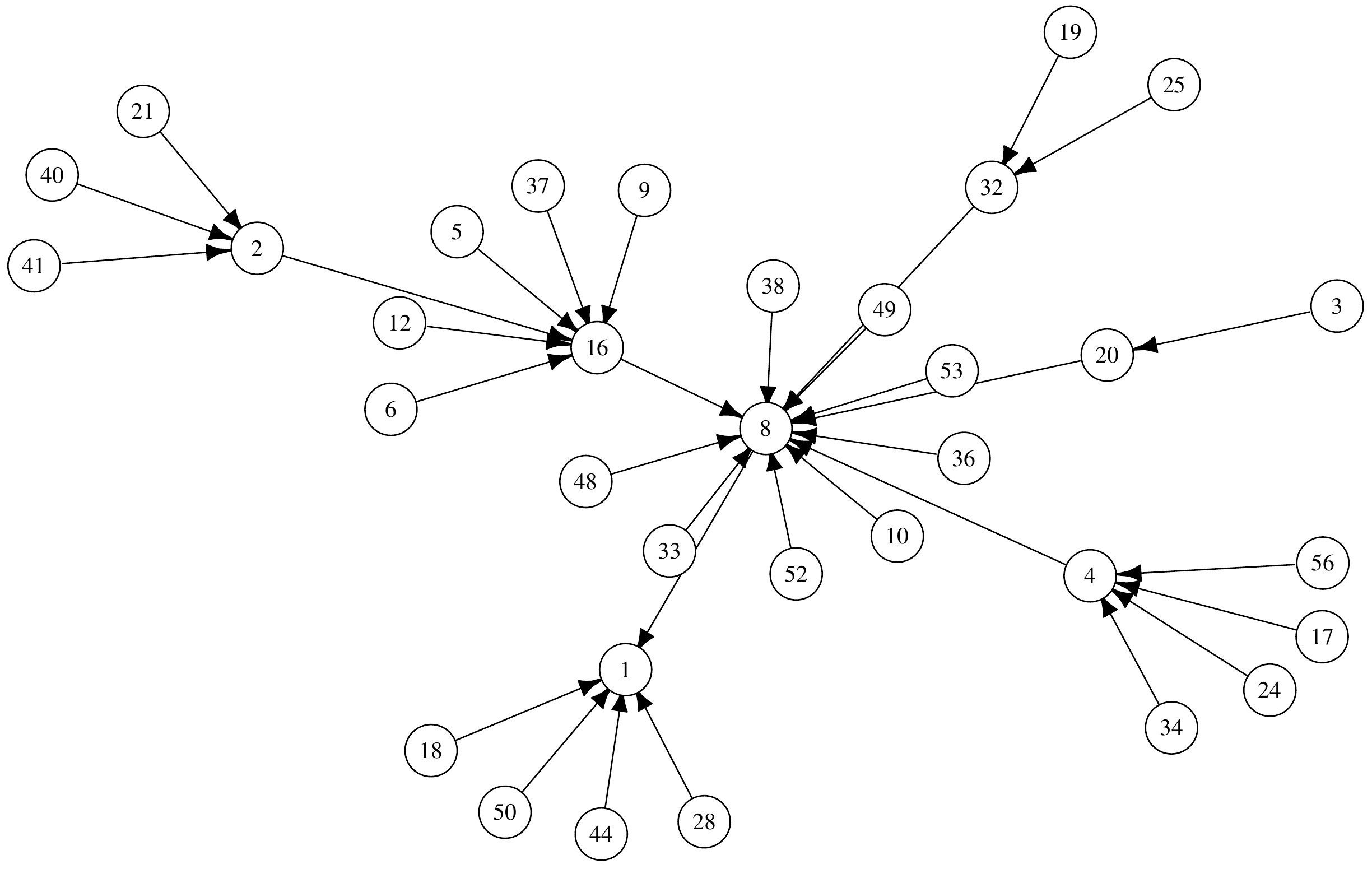}\label{fig:MaxTransitions}}
    \caption{(a)~Global graph of ${\mathcal M}$ state transitions. Edge weights are visualised by colours: from lowest weight in orange to highest weight in blue. (b)~Pruned global graph of ${\mathcal M}$:  only transitions with maximum weight for any given predecessors are shown, each node/state has at most one outgoing edge.}
    \label{fig:GlobalstateTransition}
\end{figure}       

Global transition graphs of ${\mathcal M}$ for selected inputs are shown in Fig.~\ref{fig:stateTransition}. Nodes of the graphs are states of ${\mathcal M}$, edges show transitions between the states. These directed graphs are defined as follows. There is an edge from node $a$ to node $b$ if there is such $1 \leq t \leq 1000$ that $\mathcal{M}^t=a$ and $\mathcal{M}^{t+1}=b$.  

Let us now define a weighted global transition graph $\mathcal{G}=\langle  \mathbf{Q}, \mathbf{E}, w \rangle$, where $\mathbf{Q}$ is a set of nodes (isomorphic to the $\{ 0, 1 \}^6$), and $\mathbf{E}$ is a set of edges, and weighting function 
$w: \, \mathbf{E} \rightarrow [0,1]$ assigning a number of a unit interval to each edge.  Let $a, b \in \mathbf{Q}$ and $e(a,b) \in \mathbf{E}$ then a normalised weight is calculated as $w(e(a,b))=
\frac{
\sum_{i \in \mathbf{Q}, t \in \mathbf{T}} \chi(s^t=a \text{ and } s^{t+1}=b)
}
{
\sum_{d \in \mathbf{Q}, t \in \mathbf{T}}  \sum_{\mathbf{Q}, t \in \mathbf{T}} \chi(s^t=a \text{ and } s^{t+1}=d)
}$, with $\chi$ takes value `1' when the conditions are true and `0' otherwise. In words, $w(e(a,b))$ is a number of transitions from $a$ to $b$ observed in the evolution of $\mathcal{M}$ for all possible inputs from $\mathcal{Q}$ during time interval $\bf T$ normalised by a total number of transition from $a$ to all other nodes. The graph $\mathcal{G}$ is visualised in Fig.~\ref{fig:globalGraph}. Nodes which have predecessors are  
1--6, 8--10, 12, 16--21, 24, 25, 28, 32--34, 36--38, 40, 41, 44, 48--50, 52, 53, 56. Nodes without predecessors are 7, 11, 13--15, 22, 23, 26, 27, 29--31, 35, 39, 42, 43, 45--47, 51, 54, 55, 57--63. 

Let us convert $\mathcal{G}$ to an acyclic non-weighted graph of more likely transitions $\mathcal{G}^* \langle \mathbf{Q}, \mathbf{E}^*  \rangle$, where $e(a,b) \in \mathbf{E}^*$ if $w(e(a,b))=max\{ w(e(a,c)) | e(a,c) \in \mathbf{E} \}$. That is for each node we select an outgoing edge with maximum weight. The graph is a tree, see Fig.~\ref{fig:MaxTransitions}.  Most states apart of 1, 2, 4, 8, 16, 20, 32 are Garden-of-Eden configurations, which have no predecessors. Indegrees $\nu()$ of not-Garden-of-Eden nodes are 
$\nu(20)=1, \nu(32)=2, \nu(2)=3, \nu(4)=4, \nu(1)=5, \nu(16)=6, \nu(8)=12$.
There is one fixed point, the state 1, corresponding to the situation when a spike is recorded only on electrode $e_5$; it has no successors. 

\begin{table}[]
    \centering
\begin{scriptsize}
    \begin{tabular}{c|c|l}
$t$ & $\mu(t)$ & $\mathbf{P}(t)$ \\ \hline
1 & 3 & 8, 9, 1, \\
2 & 3 & 16, 32, 8, \\
3 & 3 & 1, 16, 32, \\
4 & 3 & 8, 1, 16, \\
5 & 3 & 1, 8, 16, \\
6 & 3 & 16, 8, 1, \\
7 & 4 & 8, 1, 16, 4, \\
8 & 4 & 1, 16, 8, 5, \\
9 & 5 & 16, 1, 8, 4, 5, \\
10 & 4 & 16, 1, 8, 4, \\
11 & 5 & 8, 1, 16, 20, 4, \\
12 & 4 & 1, 16, 8, 20, \\
13 & 6 & 16, 8, 1, 17, 4, 20, \\
14 & 8 & 8, 16, 17, 4, 20, 1, 32, 2, \\
15 & 8 & 1, 16, 8, 4, 2, 10, 20, 32, \\
16 & 6 & 16, 4, 8, 1, 10, 32, \\
17 & 5 & 16, 1, 4, 8, 9, \\
18 & 7 & 8, 16, 4, 1, 17, 10, 9, \\
19 & 6 & 1, 8, 16, 17, 4, 10, \\
20 & 8 & 16, 1, 8, 17, 4, 24, 10, 2, \\
21 & 9 & 8, 16, 1, 17, 32, 24, 9, 4, 10, \\
22 & 6 & 16, 1, 8, 32, 9, 4, \\
23 & 7 & 8, 1, 16, 4, 32, 9, 17, \\
24 & 6 & 1, 16, 17, 4, 32, 8, \\
25 & 7 & 16, 1, 8, 4, 17, 32, 9, \\
26 & 6 & 8, 16, 4, 12, 1, 17, \\
27 & 6 & 1, 8, 16, 4, 17, 32, \\
28 & 6 & 16, 8, 4, 1, 24, 32, \\
29 & 7 & 8, 1, 4, 16, 12, 24, 32, \\
30 & 7 & 16, 1, 8, 4, 17, 2, 32, \\
31 & 9 & 8, 1, 24, 16, 12, 4, 2, 17, 32, \\
32 & 7 & 1, 16, 8, 24, 17, 2, 40, \\
33 & 9 & 16, 8, 1, 4, 40, 17, 24, 32, 2, \\
34 & 7 & 8, 1, 16, 24, 40, 4, 32, \\
35 & 6 & 1, 16, 8, 4, 24, 2, \\
36 & 6 & 16, 8, 1, 17, 4, 32, \\
37 & 7 & 8, 16, 17, 4, 1, 40, 2, \\
38 & 7 & 1, 8, 16, 17, 4, 24, 2, \\
39 & 7 & 16, 1, 8, 17, 9, 4, 2, \\
40 & 7 & 8, 16, 4, 1, 24, 40, 2, \\
41 & 10 & 1, 8, 16, 9, 17, 4, 18, 24, 40, 2, \\
42 & 8 & 16, 1, 8, 4, 18, 33, 40, 24, \\
43 & 9 & 8, 1, 16, 4, 24, 33, 18, 32, 34, \\
44 & 9 & 1, 16, 8, 4, 17, 33, 24, 32, 40, \\
45 & 7 & 16, 8, 4, 1, 12, 24, 34, \\
46 & 7 & 8, 1, 16, 4, 24, 18, 34, \\
47 & 5 & 1, 16, 8, 4, 33, \\
48 & 5 & 16, 8, 1, 4, 17, \\
49 & 8 & 8, 1, 16, 4, 20, 32, 24, 19, \\
50 & 6 & 1, 16, 8, 4, 17, 32, \\
51 & 8 & 16, 8, 1, 4, 17, 32, 41, 19, \\
52 & 9 & 8, 16, 4, 1, 32, 33, 41, 2, 19, \\
53 & 10 & 1, 8, 16, 4, 20, 10, 2, 41, 32, 19, \\
54 & 9 & 16, 1, 8, 5, 17, 4, 2, 32, 19, \\
    \end{tabular}
\end{scriptsize}
    \caption{Fifty four state transitions of $\mathcal{M}$ over all possible inputs: $t$ is a transition step, $\mu(t)$ is a number of different states appeared over all possible inputs, $\mathbf{P}(t)$ is a set of nodes appeared at $t$.}
    \label{tab:richnessofresponse}
\end{table}

\begin{figure}[!tbp]
    \centering
    \includegraphics[width=0.8\textwidth]{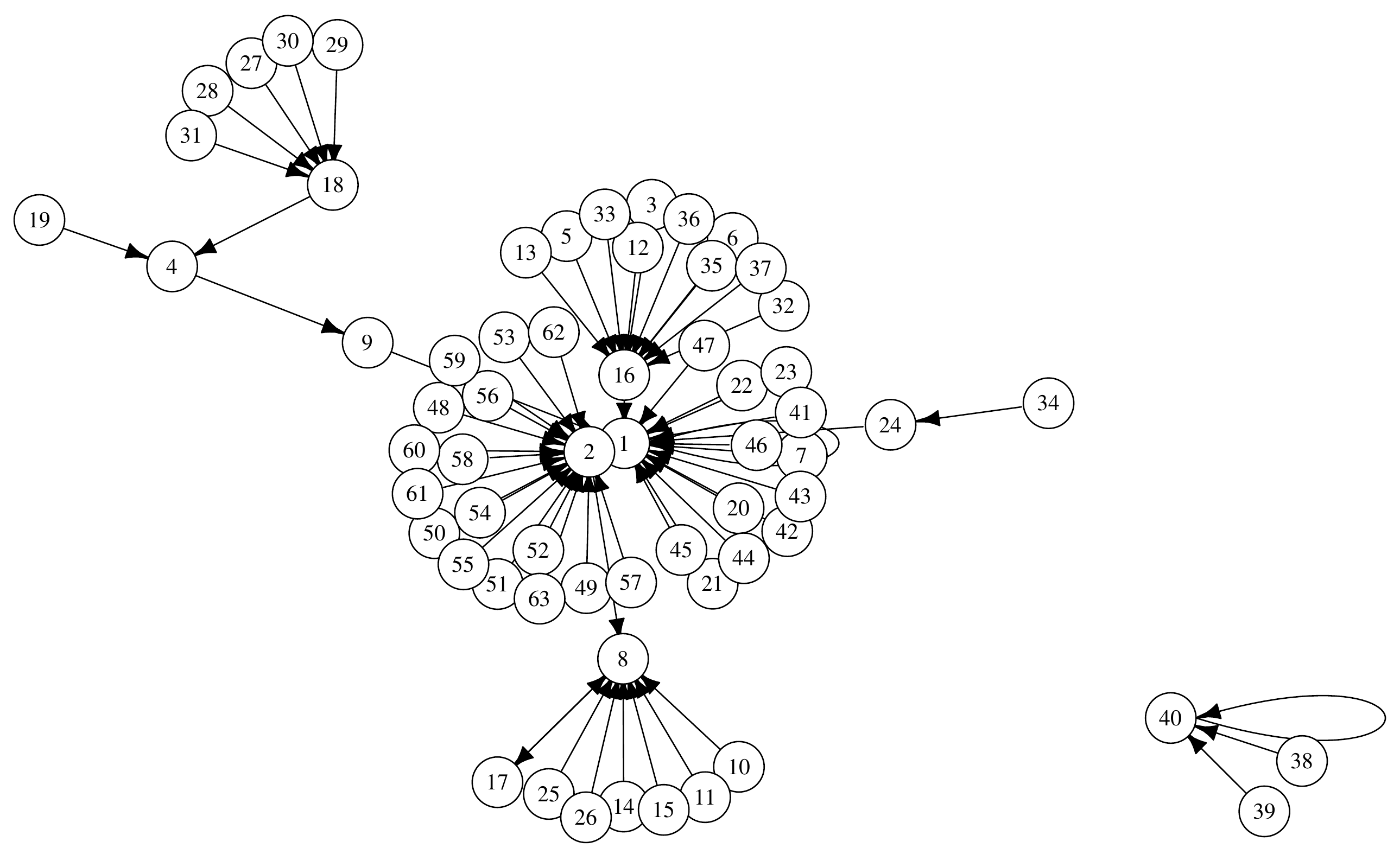}
    \caption{Graph of $g$ at $t=41$. }
    \label{fig:graphat41}
\end{figure}

  \begin{figure}[!tbp]
    \centering     
\subfigure[]{\includegraphics[width=0.48\textwidth]{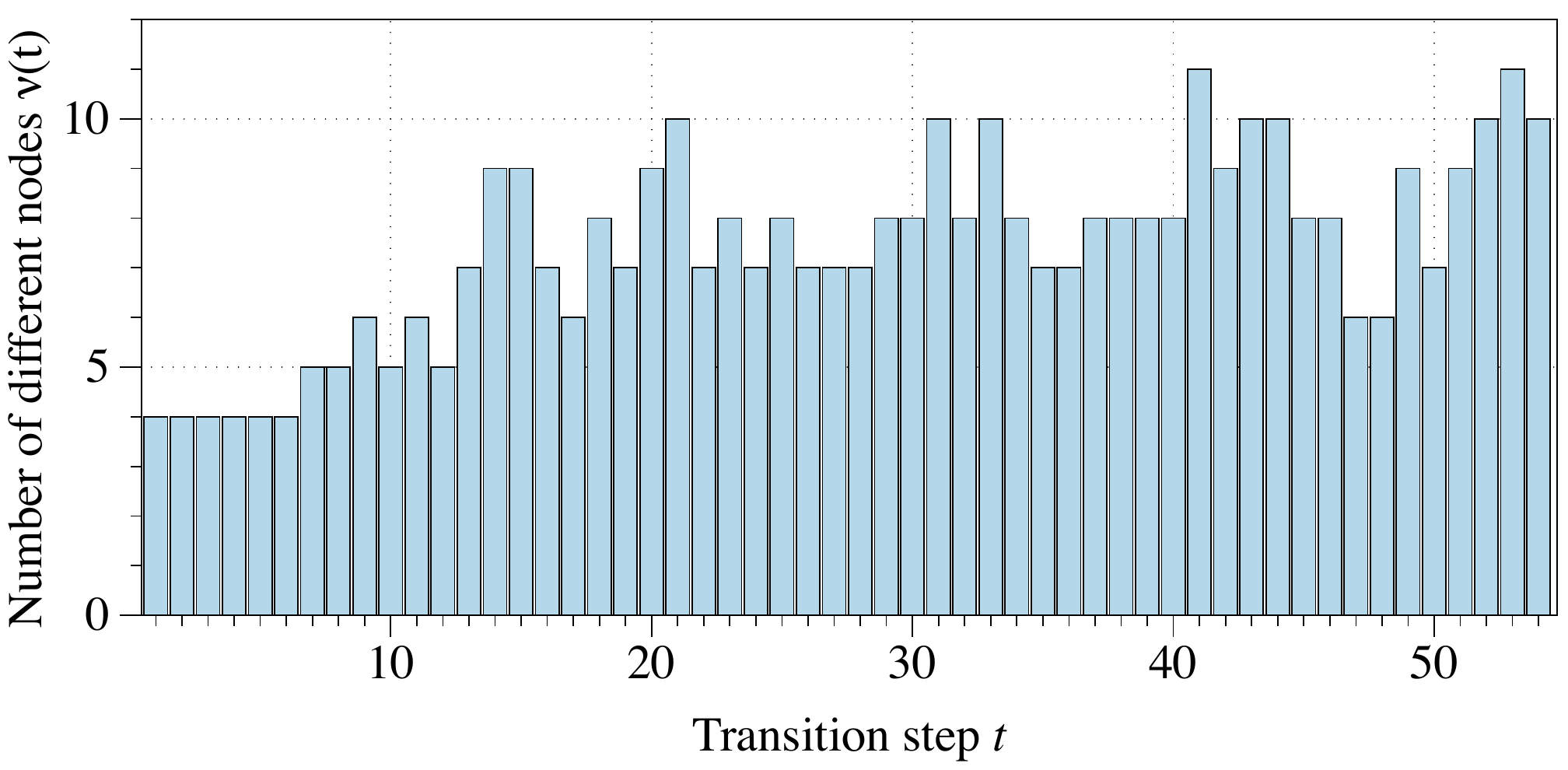}\label{fig:DifferentNodesPerTransitionOverAllInputs}}
\subfigure[]{\includegraphics[width=0.48\textwidth]{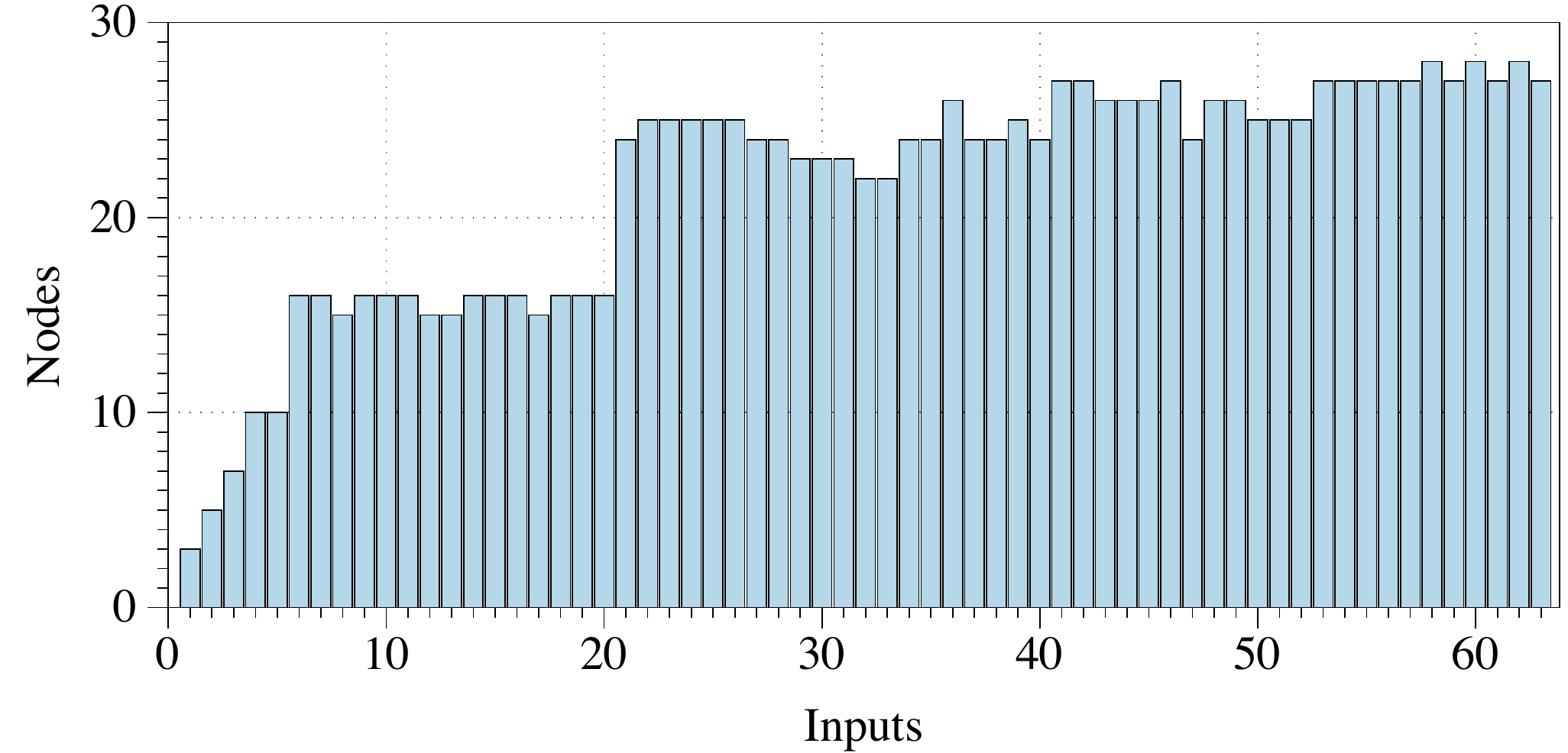}\label{fig:NodesPerInput}}
\subfigure[]{\includegraphics[width=0.48\textwidth]{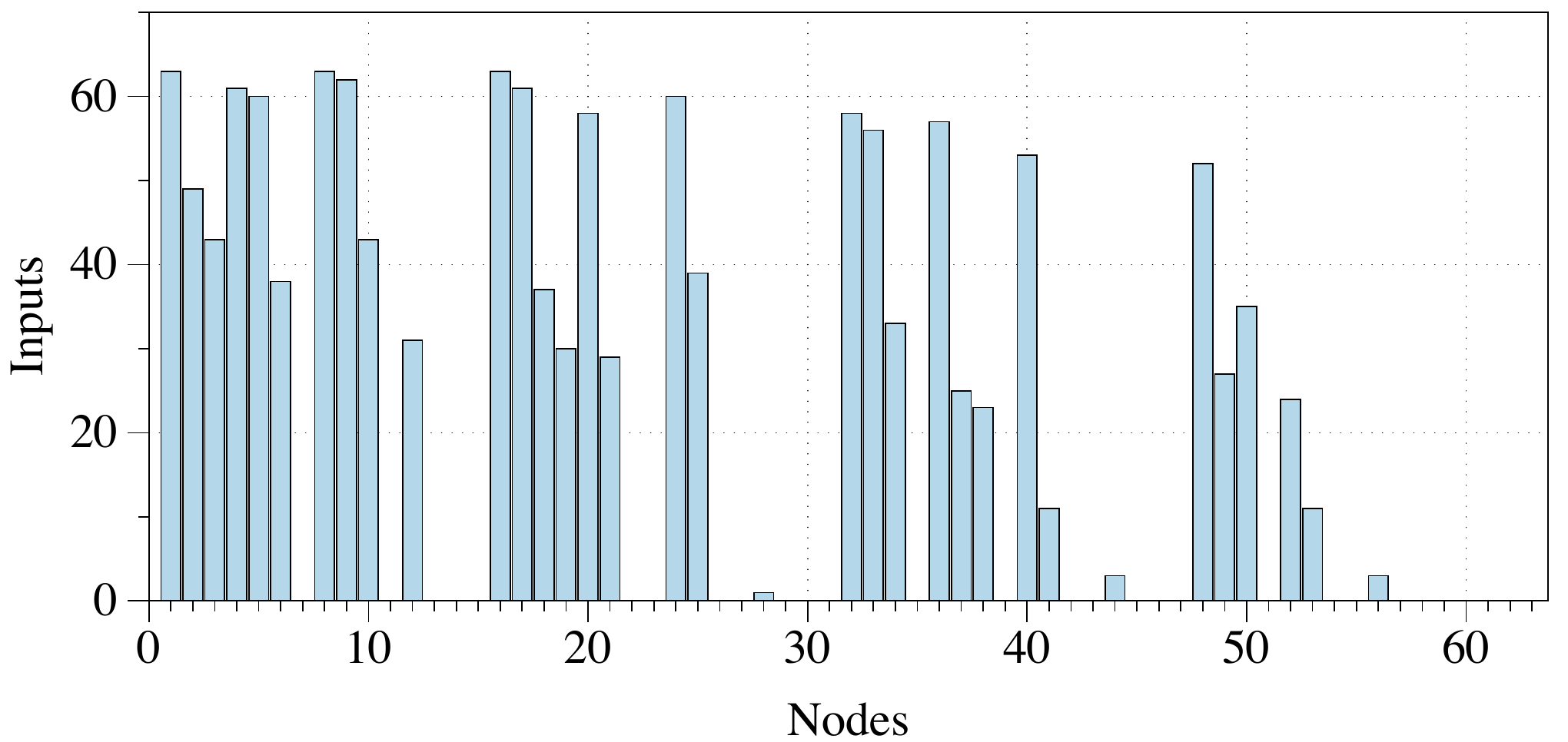}\label{fig:InputsPerNode}}
\caption{Distributions characterising richness of $\mathcal{M}$'s responses.
(a)~Different states per transitions over all inputs. Horizontal axis shows steps of $\mathcal{M}$ transitions. Vertical axis is a number of different states.
(b)~Nodes per input. Horizontal axis shows decimal values of input strings. Horizon axis shows a number of different states/nodes generates in the evolution of $\mathcal{M}$. 
(c)~Inputs per node. }
\label{fig:distributions}
\end{figure}

By analysing $\mathcal{G}$ we can characterise a richness of $\mathcal{M}$'s responses to input stimuli. We define a richness as a number of different states over all inputs, as shown in Tab.~\ref{tab:richnessofresponse}, and distribution in Fig.~\ref{fig:DifferentNodesPerTransitionOverAllInputs}. A number of states produced increases from under five for beginning of $\mathcal{M}$ evolution and then reaches circa seven states in average. Oscillations around this value are seen in (Fig.~\ref{fig:DifferentNodesPerTransitionOverAllInputs}). Figure~\ref{fig:NodesPerInput} shows a number of different nodes, generated in evolution of  $\mathcal{M}$, stimulated by a given input. There is below fifteen different states found in the evolution in responses to inputs 1 to 21 (21 corresponds to binary input string 010101); then a number of different nodes stay around 25. The diagram Fig.~\ref{fig:InputsPerNode} shows how many inputs might lead to a given state/node of $\mathcal{M}$. Some of the states/nodes are seen to be Garden-of-Eden configurations $\mathbf{E}$ (nodes without predecessors) and thus could not be generated by stimulating $\mathcal{M}$ by sequences from $\mathbf{Q}-\mathbf{E}$.

Assume $\mathbf{T}$ is a set of temporal moments when the machine responded at least to one input string with a non-zero state. Configurations at each transition $t$ can be considered as outputs representing the function $g: {0,1}^6 \rightarrow {0,1}^6$. As we can see in Tab.~\ref{tab:richnessofresponse}, transitions at $t=41$ and $t=53$ correspond to the highest number of different binary strings $(e_1, \ldots, e_6)$. The graph corresponding to $g(41)$ at $t=41$ is shown in Fig.~\ref{fig:graphat41} and is not connected. The small component consists of fixed point 40 (string `101000') with two leafs 39 (`100111') and 38 (`100110'). The largest component has a tree structure at large, with cycle 2 (`000010') -- 1 (`000001') as a root. Other nodes with most predecessors are 8 (`001000'), 16 (`010000'), and 18 (`010010').

From the transitions $g(41)$ we can reconstruct Boolean functions realised at each of six electrodes (the functions are minimised and represented in a disjunctive normal form): 
\begin{itemize}
\item[$e_0$]: $f_0(x_0, \ldots, x_5)=
 x_0 \cdot \overline{x_1 } \cdot x_2 \cdot x_3 + \overline{x_0 } \cdot x_1 \cdot \overline{x_3 } \cdot \overline{x_4 } \cdot \overline{x_5 } + x_0 \cdot \overline{x_1 } \cdot x_2 \cdot \overline{x_3 } \cdot x_4 + \overline{x_0 } \cdot \overline{x_2 } \cdot x_3 \cdot x_4 \cdot x_5 + \overline{x_0 } \cdot x_1 \cdot \overline{x_2 } \cdot x_3 \cdot \overline{x_4 } + \overline{x_1 } \cdot x_2 \cdot \overline{x_3 } \cdot \overline{x_4 } \cdot x_5 + \overline{x_0 } \cdot \overline{x_1 } \cdot \overline{x_2 } \cdot \overline{x_3 } \cdot \overline{x_4 } \cdot x_5 + \overline{x_0 } \cdot x_1 \cdot \overline{x_2 } \cdot x_3 \cdot x_4 \cdot \overline{x_5}$

\item[$e_1$]: $f_1(x_0, \ldots, x_5)=
x_0 \cdot \overline{x_1 } \cdot x_2 \cdot x_3 + \overline{x_0 } \cdot x_1 \cdot \overline{x_3 } \cdot \overline{x_4 } \cdot \overline{x_5 } + x_0 \cdot \overline{x_1 } \cdot x_2 \cdot \overline{x_3 } \cdot x_4 + \overline{x_0 } \cdot \overline{x_2 } \cdot x_3 \cdot x_4 \cdot x_5 + \overline{x_0 } \cdot x_1 \cdot \overline{x_2 } \cdot x_3 \cdot \overline{x_4 } + \overline{x_1 } \cdot x_2 \cdot \overline{x_3 } \cdot \overline{x_4 } \cdot x_5 + \overline{x_0 } \cdot \overline{x_1 } \cdot \overline{x_2 } \cdot \overline{x_3 } \cdot \overline{x_4 } \cdot x_5 + \overline{x_0 } \cdot x_1 \cdot \overline{x_2 } \cdot x_3 \cdot x_4 \cdot \overline{x_5}
$

\item[$e_2$]: $f_2(x_0, \ldots, x_5)=
x_0 \cdot \overline{x_1 } \cdot x_2 \cdot x_3 + \overline{x_0 } \cdot x_1 \cdot \overline{x_3 } \cdot \overline{x_4 } \cdot \overline{x_5 } + x_0 \cdot \overline{x_1 } \cdot x_2 \cdot \overline{x_3 } \cdot x_4 + \overline{x_0 } \cdot \overline{x_2 } \cdot x_3 \cdot x_4 \cdot x_5 + \overline{x_0 } \cdot x_1 \cdot \overline{x_2 } \cdot x_3 \cdot \overline{x_4 } + \overline{x_1 } \cdot x_2 \cdot \overline{x_3 } \cdot \overline{x_4 } \cdot x_5 + \overline{x_0 } \cdot \overline{x_1 } \cdot \overline{x_2 } \cdot \overline{x_3 } \cdot \overline{x_4 } \cdot x_5 + \overline{x_0 } \cdot x_1 \cdot \overline{x_2 } \cdot x_3 \cdot x_4 \cdot \overline{x_5}
$

\item[$e_3$]: $f_3(x_0, \ldots, x_5)=
x_0 \cdot \overline{x_1 } \cdot x_2 \cdot x_3 + \overline{x_0 } \cdot x_1 \cdot \overline{x_3 } \cdot \overline{x_4 } \cdot \overline{x_5 } + x_0 \cdot \overline{x_1 } \cdot x_2 \cdot \overline{x_3 } \cdot x_4 + \overline{x_0 } \cdot \overline{x_2 } \cdot x_3 \cdot x_4 \cdot x_5 + \overline{x_0 } \cdot x_1 \cdot \overline{x_2 } \cdot x_3 \cdot \overline{x_4 } + \overline{x_1 } \cdot x_2 \cdot \overline{x_3 } \cdot \overline{x_4 } \cdot x_5 + \overline{x_0 } \cdot \overline{x_1 } \cdot \overline{x_2 } \cdot \overline{x_3 } \cdot \overline{x_4 } \cdot x_5 + \overline{x_0 } \cdot x_1 \cdot \overline{x_2 } \cdot x_3 \cdot x_4 \cdot \overline{x_5}
$
 
\item[$e_4$]: $f_4(x_0, \ldots, x_5)=
x_0 \cdot \overline{x_1 } \cdot x_2 \cdot x_3 + \overline{x_0 } \cdot x_1 \cdot \overline{x_3 } \cdot \overline{x_4 } \cdot \overline{x_5 } + x_0 \cdot \overline{x_1 } \cdot x_2 \cdot \overline{x_3 } \cdot x_4 + \overline{x_0 } \cdot \overline{x_2 } \cdot x_3 \cdot x_4 \cdot x_5 + \overline{x_0 } \cdot x_1 \cdot \overline{x_2 } \cdot x_3 \cdot \overline{x_4 } + \overline{x_1 } \cdot x_2 \cdot \overline{x_3 } \cdot \overline{x_4 } \cdot x_5 + \overline{x_0 } \cdot \overline{x_1 } \cdot \overline{x_2 } \cdot \overline{x_3 } \cdot \overline{x_4 } \cdot x_5 + \overline{x_0 } \cdot x_1 \cdot \overline{x_2 } \cdot x_3 \cdot x_4 \cdot \overline{x_5}
$

\item[$e_5$]: $f_5(x_0, \ldots, x_5)=
x_0 \cdot \overline{x_1 } \cdot x_2 \cdot x_3 + \overline{x_0 } \cdot x_1 \cdot \overline{x_3 } \cdot \overline{x_4 } \cdot \overline{x_5 } + x_0 \cdot \overline{x_1 } \cdot x_2 \cdot \overline{x_3 } \cdot x_4 + \overline{x_0 } \cdot \overline{x_2 } \cdot x_3 \cdot x_4 \cdot x_5 + \overline{x_0 } \cdot x_1 \cdot \overline{x_2 } \cdot x_3 \cdot \overline{x_4 } + \overline{x_1 } \cdot x_2 \cdot \overline{x_3 } \cdot \overline{x_4 } \cdot x_5 + \overline{x_0 } \cdot \overline{x_1 } \cdot \overline{x_2 } \cdot \overline{x_3 } \cdot \overline{x_4 } \cdot x_5 + \overline{x_0 } \cdot x_1 \cdot \overline{x_2 } \cdot x_3 \cdot x_4 \cdot \overline{x_5}
$
\end{itemize}

\section{Discussion}
\label{discussion}

Early concepts of sub-cellular computing on cytoskeleton networks as microtubule automata~\cite{hameroff1989information,rasmussen1990computational,hameroff1990microtubule} and information processing in actin-tubulin networks~\cite{priel2006dendritic} did not specify what type of `computation' or `information processing' the cytoskeleton networks could execute and how exactly they do this. We implemented several concrete implementations of logical gates and functions on  a single actin filament~\cite{siccardi2016boolean} and  on an intersection of several actin filaments~\cite{siccardi2016logical} via collisions between solitons. We also used a reservoir-computing-like approach to discover functions on a single actin unit~\cite{adamatzky2017logical} and filament~\cite{adamatzky2018discovering}. Later, we realised that it might be unrealistic to expect someone to initiate and record a  travelling localisations (solitons, impulses) on a single actin filament. Therefore, we developed a numerical model of spikes propagating on a network of actin filament bundles and demonstrated that such a network can implement Boolean gates~\cite{AdamatzkyHuberSchnauss}. 

In present paper, we reconsidered the whole idea of the information processing on actin networks and designed an actin droplet machine. The machine is a model of a three-dimensional network, based on an experimental network developed in a droplet, which executes mapping $F$ of a space of binary strings of length $k$ on itself. The machine acts as a finite state machine, which behaviour at a low level is governed by localisations travelling along the networks and interacting with each other. By focusing on a single element of a string, i.e. a single location of an electrode, we can reconstruct $k$ functions with $k$ arguments, as we have exemplified at the end of the Sect.~\ref{machine}. Exact structure of each $k$-ary function is determined by $F$, which, in turn, is determined by the exact architecture of a three-dimensional actin network and a configuration of  electrodes. 

Thus, potential future directions could be in detailed analysis of possible architectures of actin networks developed in laboratory experiments and evaluation on how far an exact configuration of electrodes  affects  a structure of mapping $F$ and corresponding distribution of functions implementable by the actin droplet machine. The ultimate goal would be to implement actin droplet machines in laboratory experiments and to cascade several machines into a multi-processors computing architecture.

\bibliographystyle{plain}
\bibliography{bibliography}

\begin{thebibliography}{10}

\bibitem{badamatzky}
A.~Adamatzky.
\newblock Collision-based computing in biopolymers and their automata models.
\newblock {\em International Journal of Modern Physics C}, 11:1321--1346, 2000.

\bibitem{adamatzkyCBC}
Andrew Adamatzky, editor.
\newblock {\em Collision-based Computing}.
\newblock Springer, 2002.

\bibitem{adamatzky2017logical}
Andrew Adamatzky.
\newblock Logical gates in actin monomer.
\newblock {\em Scientific reports}, 7(1):11755, 2017.

\bibitem{adamatzky2018discovering}
Andrew Adamatzky.
\newblock On discovering functions in actin filament automata.
\newblock {\em arXiv preprint arXiv:1807.06352}, 2018.

\bibitem{AdamatzkyHuberSchnauss}
Andrew Adamatzky, Florian Huber, and J{\"o}rg Schnau{\ss}.
\newblock Computing on actin filament bundles.
\newblock {\em arXiv preprint arXiv:1903.10186}, 2019.

\bibitem{atienza2005probabilistic}
Felipe~Alonso Atienza, Jes{\'u}s~Requena Carri{\'o}n, Arcadi~Garc{\'\i}a
  Alberola, Jos{\'e} L~Rojo {\'A}lvarez, Juan J~S{\'a}nchez Mu{\~n}oz,
  Juan~Mart{\'\i}nez S{\'a}nchez, and Mariano~Vald{\'e}s Ch{\'a}varri.
\newblock A probabilistic model of cardiac electrical activity based on a
  cellular automata system.
\newblock {\em Revista Espa{\~n}ola de Cardiolog{\'\i}a (English Edition)},
  58(1):41--47, 2005.

\bibitem{bartles2000parallel}
James~R Bartles.
\newblock Parallel actin bundles and their multiple actin-bundling proteins.
\newblock {\em Current opinion in cell biology}, 12(1):72--78, 2000.

\bibitem{blair2002gated}
Steve Blair and Kelvin Wagner.
\newblock Gated logic with optical solitons.
\newblock In {\em Collision-based computing}, pages 355--380. Springer, 2002.

\bibitem{dowle1997fast}
Matthew Dowle, Rolf Martin~Mantel, and Dwight Barkley.
\newblock Fast simulations of waves in three-dimensional excitable media.
\newblock {\em International Journal of Bifurcation and Chaos},
  7(11):2529--2545, 1997.

\bibitem{gerhardt1990cellular}
Martin Gerhardt, Heike Schuster, and John~J Tyson.
\newblock A cellular automation model of excitable media including curvature
  and dispersion.
\newblock {\em Science}, 247(4950):1563--1566, 1990.

\bibitem{hameroff1990microtubule}
Stuart Hameroff and Steen Rasmussen.
\newblock Microtubule automata: Sub-neural information processing in biological
  neural networks, 1990.

\bibitem{hameroff1989information}
Stuart~R Hameroff and Steen Rasmussen.
\newblock Information processing in microtubules: Biomolecular automata and
  nanocomputers.
\newblock In {\em Molecular Electronics}, pages 243--257. Springer, 1989.

\bibitem{huber2013advances}
Florian Huber, J{\"o}rg Schnau{\ss}, Susanne Rönicke, Philipp Rauch, Karla
  Müller, Claus Fütterer, and Josef~A. K{\"a}s.
\newblock Emergent complexity of the cytoskeleton: from single filaments to
  tissue.
\newblock {\em Advances in Physics}, 62(1):1--112, 2013.

\bibitem{huber2012counterion}
Florian Huber, Dan Strehle, and Josef K{\"a}s.
\newblock Counterion-induced formation of regular actin bundle networks.
\newblock {\em Soft Matter}, 8(4):931--936, 2012.

\bibitem{huber2015formation}
Florian Huber, Dan Strehle, J{\"o}rg Schnau{\ss}, and Josef K{\"a}s.
\newblock Formation of regularly spaced networks as a general feature of actin
  bundle condensation by entropic forces.
\newblock {\em New Journal of Physics}, 17(4):043029, 2015.

\bibitem{jakubowski2002computing}
Mariusz~H Jakubowski, Ken Steiglitz, and Richard Squier.
\newblock Computing with solitons: a review and prospectus.
\newblock In Andrew Adamatzky, editor, {\em Collision-based computing}, pages
  277--297. Springer, 2002.

\bibitem{kavitha2017localized}
L~Kavitha, E~Parasuraman, A~Muniyappan, D~Gopi, and S~Zdravkovi{\'c}.
\newblock Localized discrete breather modes in neuronal microtubules.
\newblock {\em Nonlinear Dynamics}, 88(3):2013--2033, 2017.

\bibitem{korn1982actin}
Edward~D Korn.
\newblock Actin polymerization and its regulation by proteins from nonmuscle
  cells.
\newblock {\em Physiological Reviews}, 62(2):672--737, 1982.

\bibitem{korn1987actin}
Edward~D Korn, Marie-France Carlier, and Dominique Pantaloni.
\newblock Actin polymerization and atp hydrolysis.
\newblock {\em Science}, 238(4827):638--644, 1987.

\bibitem{lechleiter1991spiral}
James Lechleiter, Steven Girard, Ernest Peralta, and David Clapham.
\newblock Spiral calcium wave propagation and annihilation in xenopus laevis
  oocytes.
\newblock {\em Science}, 252(5002):123--126, 1991.

\bibitem{markus1990isotropic}
Mario Markus and Benno Hess.
\newblock Isotropic cellular automaton for modelling excitable media.
\newblock {\em Nature}, 347(6288):56, 1990.

\bibitem{priel2006ionic}
Avner Priel, Jack~A Tuszynski, and Horacio~F Cantiello.
\newblock Ionic waves propagation along the dendritic cytoskeleton as a
  signaling mechanism.
\newblock {\em Advances in Molecular and Cell Biology}, 37:163--180, 2006.

\bibitem{priel2006dendritic}
Avner Priel, Jack~A Tuszynski, and Horacion~F Cantiello.
\newblock The dendritic cytoskeleton as a computational device: an hypothesis.
\newblock In {\em The emerging physics of consciousness}, pages 293--325.
  Springer, 2006.

\bibitem{rasmussen1990computational}
Steen Rasmussen, Hasnain Karampurwala, Rajesh Vaidyanath, Klaus~S Jensen, and
  Stuart Hameroff.
\newblock Computational connectionism within neurons: A model of cytoskeletal
  automata subserving neural networks.
\newblock {\em Physica D: Nonlinear Phenomena}, 42(1-3):428--449, 1990.

\bibitem{sataric2009nonlinear}
MV~Satari{\'c}, DI~Ili{\'c}, N~Ralevi{\'c}, and Jack~Adam Tuszynski.
\newblock A nonlinear model of ionic wave propagation along microtubules.
\newblock {\em European biophysics journal}, 38(5):637--647, 2009.

\bibitem{sataric2011ionic}
MV~Satari{\'c} and BM~Satari{\'c}.
\newblock Ionic pulses along cytoskeletal protophilaments.
\newblock In {\em Journal of Physics: Conference Series}, volume 329, page
  012009. IOP Publishing, 2011.

\bibitem{sataric2010solitonic}
MV~Satari{\'c}, D~Sekuli{\'c}, and M~{\v{Z}}ivanov.
\newblock Solitonic ionic currents along microtubules.
\newblock {\em Journal of Computational and Theoretical Nanoscience},
  7(11):2281--2290, 2010.

\bibitem{saxberg1991cellular}
Bo~EH Saxberg and Richard~J Cohen.
\newblock Cellular automata models of cardiac conduction.
\newblock In {\em Theory of heart}, pages 437--476. Springer, 1991.

\bibitem{schnaussPRL2016}
Jörg Schnauß, Tom Golde, Carsten Schuldt, B.~U.~Sebastian Schmidt, Martin
  Glaser, Dan Strehle, Tina Händler, Claus Heussinger, and Josef~A. Käs.
\newblock Transition from a linear to a harmonic potential in collective
  dynamics of a multifilament actin bundle.
\newblock {\em Phys. Rev. Lett.}, 116(10):108102, 2016.

\bibitem{schnaussreview2016}
Jörg Schnauß, Tina Händler, and Josef~A. Käs.
\newblock Semiflexible biopolymers in bundled arrangements.
\newblock {\em Polymers}, 8(8):274, 2016.

\bibitem{siccardi2016logical}
Stefano Siccardi and Andrew Adamatzky.
\newblock Logical gates implemented by solitons at the junctions between
  one-dimensional lattices.
\newblock {\em International Journal of Bifurcation and Chaos}, 26(06):1650107,
  2016.

\bibitem{siccardi2016boolean}
Stefano Siccardi, Jack~A Tuszynski, and Andrew Adamatzky.
\newblock Boolean gates on actin filaments.
\newblock {\em Physics Letters A}, 380(1-2):88--97, 2016.

\bibitem{siregar1998interactive}
P~Siregar, JP~Sinteff, N~Julen, and P~Le~Beux.
\newblock An interactive 3d anisotropic cellular automata model of the heart.
\newblock {\em Computers and Biomedical Research}, 31(5):323--347, 1998.

\bibitem{straub1943actin}
FB~Straub.
\newblock Actin, ii.
\newblock {\em Stud. Inst. Med. Chem. Univ. Szeged}, 3:23--37, 1943.

\bibitem{straub1950muscle}
FB~Straub.
\newblock Muscle.
\newblock {\em Annual review of biochemistry}, 19(1):371--388, 1950.

\bibitem{strehle2017}
Dan Strehle, Paul Mollenkopf, Martin Glaser, Tom Golde, Carsten Schuldt,
  Josef~A. K\"as, and J\"org Schnau\ss{}.
\newblock Single actin bundle rheology.
\newblock {\em Molecules}, 22(10):1804, 2017.

\bibitem{Strehle2011}
Dan Strehle, Jörg Schnauß, Claus Heussinger, Jos{\'e} Alvarado, Mark Bathe,
  Josef~A. Käs, and Brian Gentry.
\newblock Transiently crosslinked f-actin bundles.
\newblock {\em European Biophysics Journal}, 40(1):93--101, 2011.

\bibitem{szent2004early}
Andrew~G Szent-Gy{\"o}rgyi.
\newblock The early history of the biochemistry of muscle contraction.
\newblock {\em The Journal of general physiology}, 123(6):631--641, 2004.

\bibitem{toth2009simple}
Rita Toth, Christopher Stone, Ben de~Lacy~Costello, Andrew Adamatzky, and Larry
  Bull.
\newblock Simple collision-based chemical logic gates with adaptive computing.
\newblock {\em International Journal of Nanotechnology and Molecular
  Computation (IJNMC)}, 1(3):1--16, 2009.

\bibitem{tuszynski2005molecular}
JA~Tuszy{\'n}ski, JA~Brown, E~Crawford, EJ~Carpenter, MLA Nip, JM~Dixon, and
  MV~Satari{\'c}.
\newblock Molecular dynamics simulations of tubulin structure and calculations
  of electrostatic properties of microtubules.
\newblock {\em Mathematical and Computer Modelling}, 41(10):1055--1070, 2005.

\bibitem{tuszynski1995ferroelectric}
JA~Tuszy{\'n}ski, S~Hameroff, MV~Satari{\'c}, B~Trpisova, and MLA Nip.
\newblock Ferroelectric behavior in microtubule dipole lattices: implications
  for information processing, signaling and assembly/disassembly.
\newblock {\em Journal of Theoretical Biology}, 174(4):371--380, 1995.

\bibitem{tuszynski2004results}
JA~Tuszynski, T~Luchko, EJ~Carpenter, and E~Crawford.
\newblock Results of molecular dynamics computations of the structural and
  electrostatic properties of tubulin and their consequences for microtubules.
\newblock {\em Journal of Computational and Theoretical Nanoscience},
  1(4):392--397, 2004.

\bibitem{tuszynski2004ionic}
JA~Tuszy{\'n}ski, S~Portet, JM~Dixon, C~Luxford, and HF~Cantiello.
\newblock Ionic wave propagation along actin filaments.
\newblock {\em Biophysical journal}, 86(4):1890--1903, 2004.

\bibitem{tuszynski2005nonlinear}
Jack Tuszy{\'n}ski, St{\'e}phanie Portet, and John Dixon.
\newblock Nonlinear assembly kinetics and mechanical properties of biopolymers.
\newblock {\em Nonlinear Analysis: Theory, Methods \& Applications},
  63(5-7):915--925, 2005.

\bibitem{verkhovsky1995myosin}
Alexander~B Verkhovsky, Tatyana~M Svitkina, and Gary~G Borisy.
\newblock Myosin ii filament assemblies in the active lamella of fibroblasts:
  their morphogenesis and role in the formation of actin filament bundles.
\newblock {\em The Journal of cell biology}, 131(4):989--1002, 1995.

\bibitem{wang1984reorganization}
Yu-Li Wang.
\newblock Reorganization of actin filament bundles in living fibroblasts.
\newblock {\em The Journal of cell biology}, 99(4):1478--1485, 1984.

\bibitem{weimar1992diffusion}
J{\"o}rg~R Weimar, John~J Tyson, and Layne~T Watson.
\newblock Diffusion and wave propagation in cellular automaton models of
  excitable media.
\newblock {\em Physica D: Nonlinear Phenomena}, 55(3-4):309--327, 1992.

\bibitem{ye2005efficient}
Pei Ye, Emilia Entcheva, Radu Grosu, and Scott~A Smolka.
\newblock Efficient modeling of excitable cells using hybrid automata.
\newblock In {\em Proc. of CMSB}, volume~5, pages 216--227, 2005.

\bibitem{zhang}
L.~Zhang and A.~{Adamatzky}.
\newblock Collision-based implementation of a two-bit adder in excitable
  cellular automaton.
\newblock {\em Chaos, Solitons and Fractals}, 41:1191--1200, 2009.

\end{thebibliography}

\section*{Author contributions statements} 

A.A., F.H., J.S.  undertook the research and wrote the manuscript. 

\section*{Competing interests}  
The authors declare that they have no competing interests.

\end{document}